\documentclass[aps,10pt,prl,twocolumn,groupedaddress,nofootinbib,notitlepage,preprintnumbers]{revtex4-2}

\usepackage[utf8]{inputenc}
\usepackage{amsmath}
\usepackage{amssymb}
\usepackage{graphicx} 
\usepackage{bm}        
\usepackage{amssymb}   
\usepackage{mathtools}
\usepackage[dvipsnames]{xcolor}
\usepackage{hyperref}
\usepackage{orcidlink} 
\usepackage{etoolbox}
\usepackage{xspace}
\usepackage{placeins}
\usepackage{mathrsfs}  %% for \mathscr
\usepackage{tensor}
\usepackage{soul}

\NewDocumentCommand{\sotwo}{O{red}O{black}+m}
    {%
        \begingroup
        \setulcolor{#1}%
        \setul{-.5ex}{.4pt}%
        \def\SOUL@uleverysyllable{%
            \rlap{%
                \color{#2}\the\SOUL@syllable
                \SOUL@setkern\SOUL@charkern}%
            \SOUL@ulunderline{%
                \phantom{\the\SOUL@syllable}}%
        }%
        \ul{#3}%
        \endgroup
    }
\makeatother

\newcommand{\xvn}[2]{x^{#1}_{{\gamma}#2}}
\newcommand{\xve}[2]{x^{#1}_{{\rm e}#2}}
\newcommand{\xvd}[2]{x^{#1}_{{\rm r}#2}}

\newcommand{\tauf}[1]{\tau^{(#1)}_{\rm rf}}

\newcommand{\pv}[2]{p^{#1}_{{\rm r}#2}}
\newcommand{\kv}[2]{k^{#1}_{{\rm \gamma}#2}}

\newcommand{\Xve}[2]{X^{#1}_{{\rm e}#2}}
\newcommand{\Xvd}[2]{X^{#1}_{{\rm r}#2}}
\newcommand{\Xvn}[3]{X_{\gamma#3}^{(#1)#2}}

\newcommand{\gp}[2]{g^{(#1)}_{#2}}

\begin{document}

\title{Observable Gravitational Wave Strain at Second Order}

\author{Guillem Domènech${}^{1,2}$\,
\orcidlink{0000-0003-2788-884X}}
\email{guillem.domenech@itp.uni-hannover.de}
\author{Shi Pi${}^{3,4,5}$\,
\orcidlink{0000-0002-8776-8906}}
\email{shi.pi@itp.ac.cn}
\author{Ao Wang${}^{1,3,6}$\,
\orcidlink{0009-0002-6559-5212}}
\email{wangao@itp.ac.cn}

\affiliation{%
    $^{1}$Institute for Theoretical Physics, Leibniz University Hannover, Appelstraße 2, 30167 Hannover, Germany
}%
\affiliation{
    $^{2}$Max-Planck-Institut für Gravitationsphysik, Albert-Einstein-Institut, 30167 Hannover, Germany
}%
\affiliation{
    $^{3}$Institute of Theoretical Physics, Chinese Academy of Sciences, Beijing 100190, China}
\affiliation{
    $^{4}$ Center for High Energy Physics, Peking University, Beijing 100871, China}
\affiliation{
    $^{5}$ Kavli Institute for the Physics and Mathematics of the Universe (WPI), The University of Tokyo, Kashiwa, Chiba 277-8583, Japan}
\affiliation{
    $^{6}$ School of Physical Sciences, University of Chinese Academy of Sciences, Beijing 100049, China}


\begin{abstract}
There is currently no rigorous definition of gravitational wave strain at second order in cosmological perturbation theory. The usual association of gravitational waves with transverse and traceless fluctuations of the metric on spatial hypersurfaces becomes ambiguous at second order, as it inherently depends on the spacetime slicing. While this poses no practical issues in linearized gravity, it presents a fundamental problem for secondary gravitational waves, especially notorious for gravitational waves induced by primordial fluctuations. We compute, for the first time, the observable gravitational wave strain at second order, as measured by geodesic observers that emit and receive electromagnetic signals, thereby settling the debate on gauge ambiguities. Working in a gauge invariant fashion, we find that the measured gravitational wave strain coincides with the transverse-traceless components in the Newton gauge.
\end{abstract}

\maketitle    

\textit{Introduction.}--- Interest in Gravitational Waves (GWs) has grown steadily after the first detection by the LIGO and Virgo collaborations \cite{LIGOScientific:2016aoc}. Recent hints of a GW background reported by Pulsar Timing Array (PTA) collaborations (NANOGrav, EPTA, PPTA, and CPTA) \cite{NANOGrav:2023gor,NANOGrav:2023hde,EPTA:2023sfo,EPTA:2023fyk,EPTA:2023xxk,Reardon:2023gzh,Zic:2023gta,Reardon:2023zen,Xu:2023wog} further stimulated studies on the cosmic production of GWs, see, e.g., Refs.~\cite{Binetruy:2012ze,Caprini:2018mtu,Domenech:2020ers,Roshan:2024qnv,Domenech:2024rks,Bian:2025ifp}. Among possible sources, GWs induced by primordial fluctuations in the early universe \cite{Tomita:1967wkp,Matarrese:1992rp,Matarrese:1993zf,Bruni:1996im,Matarrese:1996pp,Matarrese:1997ay,Ananda:2006af,Baumann:2007zm}, so-called induced GWs, have attracted remarkable attention; see Refs.~\cite{Yuan:2021qgz,Domenech:2021ztg} for recent reviews. 

Primordial fluctuations are the seeds of Cosmic Microwave Background (CMB) anisotropies and large-scale structures in the Universe \cite{Planck:2018vyg}. The same fluctuations
induce GWs at second order, which predict a scale-invariant spectrum $\Omega_\mathrm{GW}\sim10^{-24}$ on large scales \cite{Baumann:2007zm}. It could become significant on smaller scales, 
usually related to the primordial black hole scenario \cite{Saito:2008jc,Cai:2018dig,Byrnes:2018txb,Chen:2019xse,Ren:2023yec,Dandoy:2023jot,LISACosmologyWorkingGroup:2023njw,Ozsoy:2023ryl,Domenech:2024rks,Iovino:2024tyg,Luo:2025ewp,Iovino:2025cdy} and serve as a probe of the Universe's unexplored expansion history \cite{Cai:2019cdl,Domenech:2020kqm,Domenech:2020ssp,Domenech:2020ers,Liu:2023pau}. Yet, there are lingering gauge ambiguities in theoretical predictions for induced GWs even today \cite{Bruni:1996im,Hwang:2017oxa}. 

In linearized gravity, GWs are associated with Transverse and Traceless (TT) fluctuations of the spatial metric \cite{Maggiore:2007ulw}, commonly called tensor fluctuations. While tensor fluctuations are gauge invariant at linear order, they stop being so at second order in cosmological perturbation theory \cite{Bruni:1996im,Matarrese:1996pp,Matarrese:1997ay}, leading to notorious gauge issues in induced GWs \cite{Hwang:2017oxa,Tomikawa:2019tvi,Gong:2019mui}. Solving this matter is vital for reliable theoretical predictions and of general fundamental interest in GW physics.

Previous studies found that the predicted induced GW spectrum is identical in the Newton, synchronous, and flat gauges but not in, e.g., the comoving gauge \cite{DeLuca:2019ufz,Inomata:2019yww,Yuan:2019fwv,Chang:2020iji,Domenech:2020xin,Ota:2021fdv,Yuan:2024qfz} (although dissipative effects alleviate the issue \cite{Domenech:2025bvr}). Note that the problem is not the absence of a gauge invariant formulation, see e.g. Ref.~\cite{Domenech:2017ems}, but the lack of an explicit observable. Indeed, by means of second-order gauge transformations, Ref.~\cite{Domenech:2020xin} used the Newton gauge as a benchmark to classify an approximate gauge independence. Such a choice is based on the expectation that the Newton gauge, which recovers Newtonian physics deep inside the Hubble radius, is suitable to describe observations (see also Refs.~\cite{Matarrese:1996pp,Matarrese:1997ay}). Here we provide proof.

Existing proposals to remedy the gauge issue are based on extrapolations of results from flat spacetime. For instance, Ref.~\cite{DeLuca:2019ufz} argues that the synchronous gauge is the closest to the TT gauge. Ref.~\cite{Yuan:2025seu} proposes to drop terms that do not behave as expected from propagating radiation based on the Sommerfeld boundary condition. Although these ideas provide some insight, it is unclear whether they also apply to cosmological spacetimes without explicit calculation of the measured strain.

In this Letter, we compute, for the first time, the GW strain measured by a detector at second order. For simplicity, we focus on geodesic clocks \cite{marzke1964gravitation}, where two geodesic observers exchange ElectroMagnetic (EM) signals. Time delays are observables and therefore free from gauge ambiguities. We find that the physical effect due to GWs coincides with the TT components in the Newton gauge, thereby settling the gauge issue of induced GWs. Although our analysis strictly applies to the idealized case of geodesic observers, we argue in the conclusions that generalizations to non-geodesic observers, such as in ground-based interferometers, are possible.

\textit{GW measurements and cosmic geodesic clock.}--- Current GW observations rely on EM signal exchanges \cite{Romano:2016dpx}. For example, in GW interferometers, a laser beam is emitted, split, bounced off mirrors, and received back. PTAs carefully monitor periodic radio signals emitted by millisecond pulsars in the galaxy. Similarly, one may use spacecraft Doppler tracking \cite{Estabrook:1975jtn,Armstrong:2006zz,Zwick:2024hag}. In all these measurements, passing GWs cause a delay in the proper time the detector takes to receive the signal, as well as a shift in the signal's frequency.

\begin{figure}
    \center
    \includegraphics[width=0.4\textwidth]{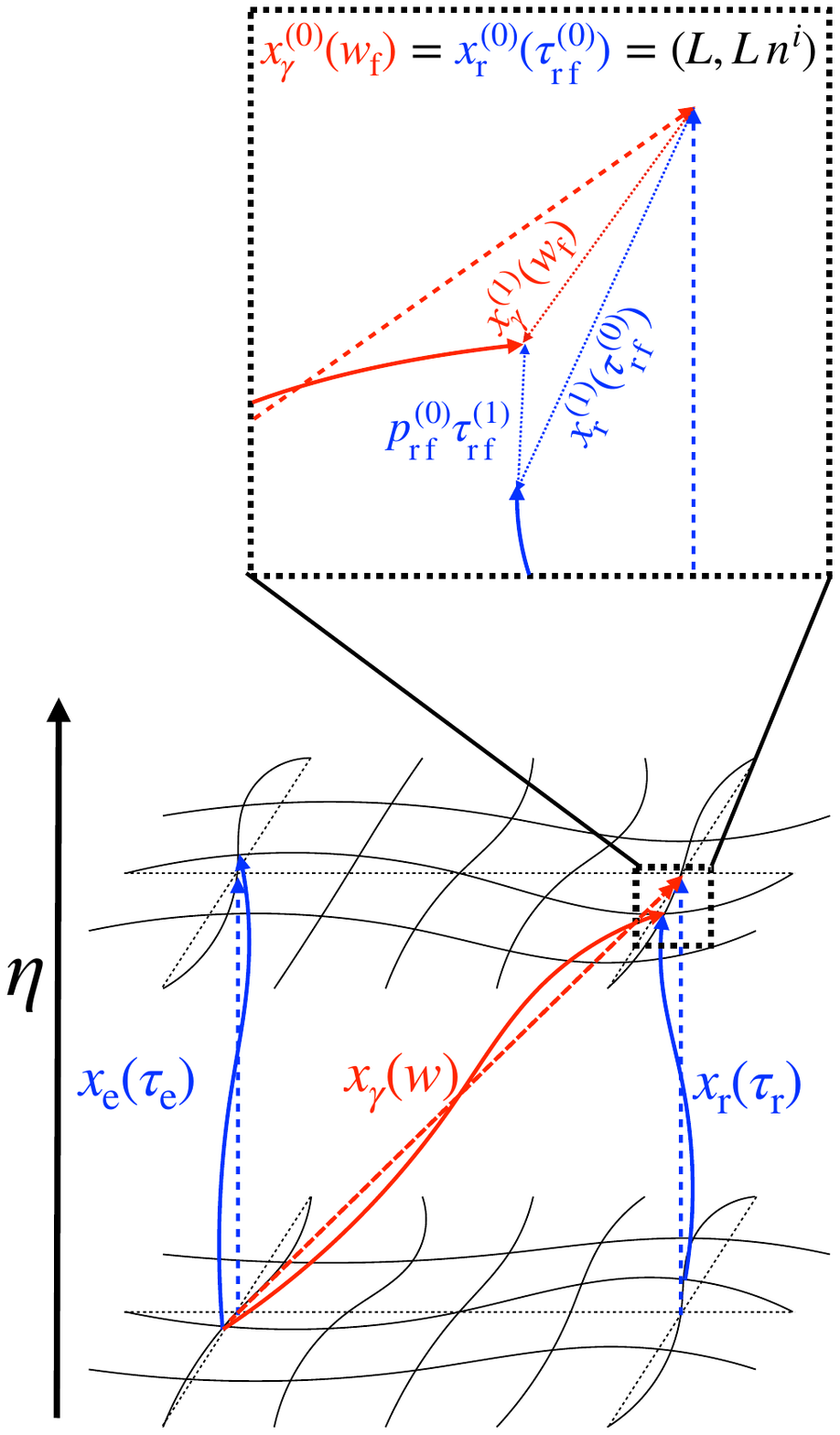}
    \caption{\textbf{[Lower]} Illustration of the elementary process. Time flows upwards. Black lines depict the time slices, blue lines show the world lines of the observers, and red lines show the propagating light rays. Dashed and solid lines, respectively, correspond to the background and perturbed trajectories. \textbf{[Upper]} Zoom-in of reception showing first-order corrections as in Eq.~\eqref{eq:fc}.
    }
    \label{fig:config}
\end{figure}

The elementary process in such measurements is the emission and reception of photons between two separate time-like observers. Let us call the spacetime trajectories of emitter and receptor respectively as $x_{\rm e}^\mu(\tau_{\rm e})$ and $x_{\rm r}^\mu(\tau_{\rm r})$, where $\tau$ is the corresponding proper time, and the photon geodesic connecting them as $x_\gamma^\mu(w)$, where $w$ is the affine parameter. We provide an illustration in Fig.~\ref{fig:config}.

A photon emitted at initial time $\tau_{\rm ei}$ arrives at the receptor at final time $\tau_{\rm rf}$. For initially synchronized observers, $\tau_{\rm rf}$ is the time delay of the signal. Note that, formally, the synchronization time slice corresponds to the constant proper time hypersurface used, e.g., in laying out large-scale structure observables in Refs.~\cite{Schmidt:2012ne,Jeong:2013psa,Jeong:2014ufa,Dai:2015rda}. Irrespective of synchronization, pulses periodically emitted every \textit{infinitessimal} time step $\Delta\tau_{\rm e}$ are received every $\Delta\tau_{\rm r}$ with redshift $\Delta\tau_{\rm r}/\Delta\tau_{\rm e}\to d\tau_{\rm rf}(\tau_{\rm ei})/d\tau_{\rm ei} =1+z$ \cite{brill_redshift}. Our goal is to compute variations of the time delay and redshift due to passing GWs at second order in cosmological perturbation theory. Here we show our main results, 
and leave a detailed derivation to a companion paper \cite{InPrep}.

To build our geodesic clock, we solve the time-like geodesics of the observers, and the null geodesic of the photons in a perturbed flat Friedmann–Lemaître–Robertson–Walker (FLRW) Universe, the latter with initial and final boundary conditions at emission and reception, \textit{i.e.} $x_{\gamma}^\mu(w_{\rm i})=x_{\rm e}^\mu(\tau_{\rm ei})$ and $x_{\gamma}^\mu(w_{\rm f})=x_{\rm r}^\mu(\tau_{\rm rf})$. %Before showing the results, 
Let us sketch our methodology below. 

We solve the system perturbatively in terms of proper reception time, which is gauge invariant by definition. Expanding the boundary condition at reception around a background proper reception time $\tauf{0}$ allows us to isolate perturbative corrections. For instance, at first order, we have that
\begin{align}
   \xvd{(1)\mu}{}(\tauf{0})+\pv{(0)\mu}{}(\tauf{0})\,\tauf{1}=\xvn{(1)\mu}{\rm f}\label{eq:fc}\,,
\end{align}
where $\pv{\mu}{}(\tau_{\rm r})\equiv d\xvd{\mu}{}/d\tau_{\rm r}$ is the receptor's 4-momentum and, here and henceforth, we use a superscript $(l)$ to denote the order in the perturbation expansion.  With the null geodesic solution, the zeroth component of Eq.~\eqref{eq:fc} links $\tauf{1}$ to $\tau_{\rm ei}$ and metric perturbations in between. We proceed in a similar fashion at the second order.

In conformal coordinates $x^\mu=(\eta,x^i)$, the perturbed FLRW metric is \cite{Domenech:2017ems}
\begin{align}
    {\rm d} s^2 =&g_{\mu\nu}dx^\mu dx^\nu = a^2(- e^{2\phi} {\rm d} \eta^2\nonumber\\&\quad\,+ e^{2\psi}\big(e^Y\big)_{ij}({\rm d} x^i+N^i {\rm d} \eta)({\rm d} x^j+N^j {\rm d} \eta)).
\end{align}
In the scalar-vector-tensor decomposition \cite{Kodama:1984ziu,Mukhanov:1990me,Malik:2008im}, and ignoring vector fluctuations for simplicity, we have that $N^i=\delta^{ij}\partial_j\beta$ and $Y_{ij}=2(\partial_i\partial_j-\delta_{ij}\partial^2/3)E+h_{ij}$, with $h_{ij}$ the TT components usually associated with GWs. Note that we exclusively use $\partial^2\equiv \delta^{ij}\partial_i\partial_j$ for the Laplacian.

To keep it general, we do not completely specify the initial time slice, but only fix the relative spacetime background locations of emitter and receiver at rest. Namely, we assume they are separated in a direction $n^i$ by a comoving distance $L$, that is we fix $\xvd{(0)\mu}{}-\xve{(0)\mu}{}  =(0,L\,n^{i})$ but leave $\xvd{(1)\mu}{}$ and $\xve{(1)\mu}{}$ unspecified. This allows us to carry out a consistent gauge invariant derivation. We leave $n^{i}$ free, as one could imagine a sphere of detectors exchanging EM signals in all directions.  Let us emphasize that proper time delays due to passing GWs (which are gauge invariant) must be independent of the specific initial conditions. For example, one may consider a situation with no initial spacetime fluctuations around the detector, only to be perturbed after emission.

\textit{First-order results.}--- Let us take a closer look at first-order corrections, since they also appear at second-order. After integrating the null condition, \textit{i.e.} $\kv{\mu}{}\kv{}{\mu}=0$ with $\kv{\mu}{}\equiv dx_\gamma^\mu/d w$ the photon's 4-momentum, we calculate the first-order time delay in \eqref{eq:fc} by integrating by parts and using boundary conditions, which reads
\begin{subequations}\label{eq:timedelayone}
\begin{align}
    {\tauf{1}}=a_{\rm rf}\Big(&\int_{0}^{L}\hspace{-0.1mm} {\rm d}\lambda\,\Big(\frac{1}{2}\,n^{i}n^{j}h_{ij}^{(1)}-\Phi^{(1)}-\Psi^{(1)} \Big)\label{eq:firstline}\\
    &-X^{(1)0}_{\rm rf}+X^{(1)0}_{\rm ei}+n_{i}(X^{(1)i}_{\rm rf}-X^{(1)i}_{\rm ei})\Big)\,,\label{eq:secondline}
\end{align}
\end{subequations}
where $\Phi^{(1)}$ and $\Psi^{(1)}$ are the (gauge invariant) Bardeen potentials \cite{Malik:2008im} and we defined ${\rm d}\lambda=dw/a^2(w)$. Fixing $\lambda_{\rm i}=0$, the background solution gives $\lambda_{\rm f}=L$. We also introduced the gauge-invariant spacetime locations,
\begin{equation}
    X^{(1)0}=x^{(1)0}-\sigma^{(1)},\quad X^{(1)i}=x^{(1)i}+E^{(1)}_{,i}\,,
\end{equation}
with $\sigma=\beta-E'$ the spacetime shear. 

The time delay \eqref{eq:timedelayone} is manifestly gauge invariant. The first line \eqref{eq:firstline} contains the well-known GW and Shapiro time delay, respectively. The second line \eqref{eq:secondline} includes the general initial conditions and the observer's motion effects. This is clearer if we split the receptor's location into initial condition and geodesic evolution, namely $x^{(1)\mu}_{\rm rf}=x^{(1)\mu}_{\rm ri}+\Delta x^{(1)\mu}_{\rm r}$, where $\Delta x^{(1)\mu}_{\rm r}=\int_{\eta_{\rm i}}^{\eta_{\rm f}} d\eta \,a p^{(1)\mu}_{\rm r}$. Doing so, we recast the second line \eqref{eq:secondline} as
\begin{subequations}\label{eq:timedelaysecondline}
\begin{align}
    {\tauf{1}}\supset a_{\rm rf}&\Big(
    -\Delta T_{\rm i}+\Delta L_{\rm i}+\Big(1-\frac{a_{\rm ri}}{a_{\rm rf}}\Big)X^{(1)0}_{\rm ri}\label{eq:timedelaysecondline1}\\&\quad+\int_{\eta_{\rm i}}^{\eta_{\rm f}}d\eta \,n^iV^{(1)}_{{\rm r},i}+\frac{1}{a_{\rm rf}}\int_{\eta_{\rm i}}^{\eta_{\rm f}}d\eta \,a \Phi^{(1)}\Big)\,,\label{eq:timedelaysecondline2}
\end{align}
\end{subequations}
where we used that $v_{{\rm r},i}=p^{(1)i}_{{\rm r}}/p^{(0)}_{{\rm r}0}$ and introduced  the (gauge invariant) velocity potential, $V= v+E'$. We also defined $\Delta T_{\rm i}=X^{(1)0}_{\rm ri}-X^{(1)0}_{\rm ei}$ and $\Delta L_{\rm i}\equiv n_i(X^{(1)i}_{\rm ri}-X^{(1)i}_{\rm ei})$ for later convenience. At the background we have that $\eta_{\rm f}-\eta_i=L$. The solution for $V_{{\rm r}}$ can be found in the \textit{End Matter}. 

Now we show that the effects of unspecified initial conditions can be absorbed into a redefinition of the background time delay. Noting that proper time is given by $d\tau_{\rm r} =d\eta/p^{0}_{\rm r}$ and so $\tau_{\rm rf}^{(0)}=\int_{ \eta_{\rm i}}^{\eta_{\rm f}}a(\eta)d\eta$, we identify that the terms in \eqref{eq:timedelaysecondline1} correspond to a redefinition of $\tau_{\rm rf}^{(0)}$ via
\begin{align}\label{eq:shiftdiscussion}
\eta_{\rm i}\to \eta_{\rm i}+X^{(1)0}_{\rm ri}\,\,\,,\,\,\,
\eta_{\rm f}\to \eta_{\rm f}+X^{(1)0}_{\rm ri}-\Delta T_{\rm i}+\Delta L_{\rm i}\,.
\end{align}
For the receptor, the signal left $\Delta T_{\rm i}$ earlier and traveled $\Delta L_{\rm i}$ more so it arrived $\Delta T_{\rm i}$ earlier and $\Delta L_{\rm i}$ later, in addition to a shift of the receptor's conformal time by $X^{(1)0}_{\rm ri}$.
The remaining terms in \eqref{eq:timedelaysecondline2} are due to the observer's motion, respectively due to the velocity and time delay gained from falling in a gravitational potential $\Phi^{(1)}$. The physical time delay fluctuation is then given by
\begin{subequations}\label{eq:timedelayone2}
\begin{align}
    \Delta\tau^{(1)}=a_{\rm rf}&\int_{0}^{L}\hspace{-0.1mm} {\rm d}\lambda\,\Big(\frac{1}{2}\,n^{i}n^{j}h_{ij}^{(1)}-\Phi^{(1)}-\Psi^{(1)} \Big)\label{eq:firstline2}\\
    &+a_{\rm rf}\int_{0}^{L}d\eta \,n^iV^{(1)}_{{\rm r},i}+\int_{0}^{L}d\eta \,a \Phi^{(1)}\,,\label{eq:secondline2}
\end{align}
\end{subequations}
where we set $\eta_i=0$. The time delay \eqref{eq:timedelayone2} coincides with the one evaluated in constant proper time slices, which corresponds to $\Delta T_{\rm i}=X^{(1)0}_{\rm ri}=0$ and a redefinition $L\to L+\Delta L_{\rm i}$, and therefore matches construction of the ``cosmic clocks'' of Ref.~\cite{Jeong:2013psa} and the Fermi normal coordinate approach of Refs.~\cite{Schmidt:2012ne,Jeong:2014ufa,Dai:2015rda}.

Despite being gauge invariant, the time delay \eqref{eq:timedelayone2} is not very useful for practical purposes without a complete knowledge of the initial state of the geodesic clock, or equivalently, a reconstruction of constant proper time slices. Although out of the scope of this paper, one could imagine a setup where the initial separation and proper motion of emitter and receiver are known by other means. If so, one could isolate the propagation effects.

In practice, it is more appropriate to consider the variations in the time delay from successive infinitesimally close pulses, namely the redshift. Using that $1+z=d\tau_{\rm rf}(\tau_{\rm ei})/d\tau_{\rm ei}$, we find that the first order correction reads
\begin{align}\label{eq:z1}
    \frac{z^{(1)}}{1+\tilde z}=&\, \Phi^{(1)}_{\rm rf}-\Phi^{(1)}_{\rm ei}- \int_0^{L} {\rm d}\lambda\,\big({\Phi^{(1)}}'+{\Psi^{(1)}}'\big)\notag\\
    &\quad\,
    +n^i(V^{(1)}_{{\rm r},i}-V^{(1)}_{{\rm e},i})+ \int_0^{L} {\rm d}\lambda\, n^in^j{h^{(1)}_{ij}}'\,,
\end{align}
where we introduced the measured redshift $1+\tilde z\equiv {a_{\rm r}(\tau_{\rm r})}/{a_{\rm e}(\tau_{\rm e}})$ \cite{Bonvin:2011bg}, with $\tau$ being the non-linear proper time of emitter and observer. Eq.~\eqref{eq:z1} coincides with the conventional definition, that is $1+z=(p_\mu k_\gamma^\mu)_{\rm r}/(p_\mu k_\gamma^\mu)_{\rm e}$, and contains the standard and integrated Sachs-Wolfe, the Doppler and the GW effects \cite{Durrer:1993tti,DiDio:2012bu}, also recovered using ``cosmic clocks'' \cite{Jeong:2013psa}.

\textit{Second-order results.}--- From Eqs.~\eqref{eq:firstline2} and \eqref{eq:z1}, we see that the GW effect comes from the term proportional to $n^in^j$.
As general second-order calculations are quite involved (see, e.g., Refs.~\cite{Gasperini:2011us,Ben-Dayan:2012lcv,Bertacca:2014wga,Marozzi:2014kua,Yoo:2017svj,Fuentes:2019nel,Magi:2022nfy,Bechaz:2025ojy}), we focus solely on terms containing $n^in^j$ from now on. These terms lead to the characteristic ``quadrupole'' nature of GWs.

We now proceed as in the first-order case. We expand the boundary condition $x_{\gamma}^\mu(w_{\rm f})=x_{\rm r}^\mu(\tau_{\rm rf})$ up to second order, for coordinates and proper time, and use the integrated null condition with the solutions to the first order null geodesics to relate $\tauf{2}$ to the initial boundary condition. We provide more details in the \textit{End Matter}.

After integration by parts, we find that the quadrupole component of the second-order time delay is given by
%\begin{align}\label{eq:stdt}
%    \tauf{2}\supset a_{\rm rf}&\,n^in^j\Bigg(
%    \frac{1}{2}\int_0^{L}{\rm d}\lambda \, h^{N(2)}_{ij}+V^{(1)}_{{\rm rf},i} \Delta X^{(1)}_{j}\notag\\
%    &\quad\qquad+\left(\frac{a'_{\rm rf}}{2a_{\rm rf}^2}-\frac{a_{\rm rf}^4}{2w_{\rm f}^2}L\right)\Delta X^{(1)}_{i} \Delta X^{(1)}_{j}\Bigg)\,,
%\end{align}
\begin{align}\label{eq:stdt}
     &\tauf{2}\supset a_{\rm rf}\,n^in^j\notag\\&\,\times\bigg(\frac{1}{2}
    \int_0^{L}{\rm d}\lambda \left(h^{(2)}_{ij}+\tensor{P}{_{ij}^{lk}}\left(\sigma^{(1)}_l\sigma^{(1)}_k+E^{(1)}_{,lm}E^{(1)}_{,mk}\right)\right)\notag\\
    &\,\,\,\,\,+V^{(1)}_{{\rm rf},i} \Delta X^{(1)}_{j}+\left(\frac{a'_{\rm rf}}{2a_{\rm rf}^2}-\frac{a_{\rm rf}^4}{2w_{\rm f}^2}L\right)\Delta X^{(1)}_{i} \Delta X^{(1)}_{j}\bigg)\,,
\end{align}
where $\tensor{P}{_{ij}^{lk}}$ is the TT projector, $\Delta X^{(1)}_{i}=\int_{0}^{L}d\eta \,V^{(1)}_{{\rm r},i}$ and we dropped terms proportional to the initial displacements from \eqref{eq:timedelaysecondline1}. Remarkably, the propagation effects (second line in Eq.~\eqref{eq:stdt}) are gauge invariant and %$h^{N(2)}_{ij}$ is the gauge invariant combination which 
coincide with the TT components in Newton gauge, that is \cite{Domenech:2017ems}
\begin{equation}\label{eq:hnij}
    h^{N(2)}_{ij}\equiv h^{(2)}_{ij}+\tensor{P}{_{ij}^{lk}}\left(\sigma^{(1)}_l\sigma^{(1)}_k+E^{(1)}_{,lm}E^{(1)}_{,mk}\right)\,,
\end{equation}
where recall that the Newton gauge corresponds to $\sigma=E=0$. We conclude that the gauge-invariant time delay due to GWs at second order is given by
\begin{align}\label{eq:stdtGW}
    \tau_{\rm GW}^{(2)} = a_{\rm rf}&\,n^in^j
    \frac{1}{2}\int_0^{L}{\rm d}\lambda \, h^{N(2)}_{ij}\,,
\end{align}
which can be evaluated in any gauge via Eq.~\eqref{eq:hnij}.

Eq.~\eqref{eq:stdtGW} is the main result of our work. It explicitly shows that the detector response to passing GWs is sensitive to the gauge invariant combination $h^{N(2)}_{ij}$. In other words, theoretical predictions for GWs are more conveniently derived in the Newton gauge, where $h_{ij}$ coincides with $h^{N(2)}_{ij}$, which is what a GW detector measures. Let us emphasize that our formalism is gauge invariant and that we did not fix any gauge. In other words, Eqs.~\eqref{eq:stdtGW} and  \eqref{eq:hnij} are valid in every gauge. It so happened that propagation effects in the time delay of a geodesic clock, which is a gauge invariant, observable quantity, are directly proportional to $h^{N(2)}_{ij}$.

The quadrupole component in the time delay \eqref{eq:stdt} also contains boundary terms proportional to $\Delta X^{(1)}_{j}$. These terms are exclusively due to the observer's motion. For instance, some of these terms are proportional to $n^in^jV_{{\rm r},i}V_{{\rm r},j}$. If we consider our sphere of receptors, such boundary terms originate from dynamical deformations of the initial spherical configuration. But they have nothing to do with passing GWs. Thus, if one keeps track of the receptor's motions by independent means, e.g., via additional exchange of signals between receptors, one can isolate the propagation effects and measure $h^{N(2)}_{ij}$. However, as in the first-order case, using the time delay requires extensive knowledge of the initial configuration.

Our results can be applied to more general situations. Again, it is intriguing to start from the first-order results. Since time delay is subject to initial subtleties, let us consider the exchange of pulses and the resulting redshift \eqref{eq:z1}. Each term in \eqref{eq:z1} exhibits a distinct pattern, which can be used to disentangle the different effects. For example, in the context of PTAs, we can first average over time and over pulsars at different positions. This average is insensitive to propagation effects. Then, one can reconstruct the gravitational potential to account for the time-dilation effects, that is, those associated with the expansion and the acceleration induced by the gravitational potential. This ideal procedure allows us to isolate the propagation effect at different times.

To compare with PTA literature, consider a coherent plane GW along the $x=x^1$ direction given by $h^{(1)}_{ij}=h(t-x) e^{+/\times}_{ij}$, where $ e^{+/\times}_{ij}$ are the $+$ or $\times$ polarization tensors. Using Eq.~\eqref{eq:z1}, we recover the well-known GW induced redshift \cite{Estabrook:1975jtn,Romano:2023zhb}, namely
\begin{equation}
    z^{(1)}_{\rm GW}=\,\frac{1}{2}\frac{n^i n^j e^{+/\times}_{ij}}{1-n^x} \big(h(t-L\, n^x)-h(t)\big)\,,
\end{equation}
where for PTAs we approximated ${a_{\rm rf}}\approx{a_{\rm ei}}$. The accumulative effect of the redshift on the pulse arrival time is called the timing residual. Note that the timing residual is equal to the time delay \eqref{eq:timedelayone} due to GWs up to a time-independent strain.
In PTAs, angular correlations of the GW-induced redshift follow the so-called Hellings–Downs angular correlation; a direct consequence of the quadrupolar nature of GWs \cite{Hellings:1983fr}.

Similarly, from the second-order time-delay \eqref{eq:stdtGW}, we conclude that the redshift induced by passing GWs at second-order is given by
\begin{align}\label{eq:z2}
    z^{(2)}_{\rm GW}&=\frac{a_{\rm rf}}{a_{\rm ei}} \,\frac{1}{2}n^in^j\int_0^{L}{\rm d}\lambda \, {h^{N(2)}_{ij}}'.
\end{align}
%\textcolor{red}{
%\begin{align}\label{eq:z22}
%    &z^{(2)}_{\rm GW}=\frac{a_{\rm rf}}{a_{\rm ei}} \,\frac{1}{2}n^in^j\int_0^{L}{\rm d}\lambda\nonumber\\&\quad\quad\,\times \frac{\partial}{\partial\eta}\left( h^{(2)}_{ij}+\tensor{P}{_{ij}^{lk}}\left(\sigma^{(1)}_l\sigma^{(1)}_k+E^{(1)}_{,lm}E^{(1)}_{,mk}\right)\right).
%\end{align}}
This shows that the second-order TT component in the Newton gauge directly maps to the GW effect. Our results demonstrate that the gauge-invariant GW strain power spectrum of $h^{N(2)}_{ij}$, given by Eq.~\eqref{eq:hnij}, is the physical, i.e., observable, quantity for induced GWs. With this, we settle the debate on the gauge issue of induced GWs.

At this point, let us discuss implications of our results on the time-independent GW strain induced by the constant gravitational potential $\Phi^{(1)}$ in matter dominated universes \cite{Mollerach:2003nq,Baumann:2007zm,Assadullahi:2009nf,Hwang:2017oxa,Domenech:2020xin,Papanikolaou:2020qtd,Sipp:2022kmb}. This constant solution, if generated in early matter dominated phases, has been shown to be very sensitive to the transition to the radiation dominated phase \cite{Inomata:2019ivs,Inomata:2019zqy}. Looking at Eq.~\eqref{eq:z2}, it is clear that this constant mode does not affect the redshift. Therefore, GW detectors using time-series data (such as PTAs and Doppler tracking) will not detect such a time-independent strain. The same applies to CMB temperature fluctuations and B-modes \cite{Naruko:2013aaa,Saito:2014bxa,Namikawa:2021obu,Gurian:2021rfv} as they also depend on the redshift. Furthermore, it does not affect the GW energy density if we consider a definition in which $\rho_{\rm GWs}\propto \langle h_{ij}'h_{ij}'\rangle$ as proposed by Refs.~\cite{Cai:2021jbi,Ota:2021fdv,Sipp:2022kmb}.

However, despite not propagating as a GW, one may wonder whether such a time-independent strain could be measured at all. In fact, such constant strain also appears in the Post-Newtonian approximation \cite{Matarrese:1996pp}. Our interpretation, based on our results of the second-order time delay \eqref{eq:stdtGW}, is that such a constant strain behaves like the GW memory effect \cite{Favata:2010zu,Harte:2024mwj}. 
 For instance, once the strain becomes constant, it would appear as 
\begin{align}\label{eq:stdt2}
    \tauf{2}\supset a_{\rm rf}\,
    &\frac{1}{2}n^in^j\int_0^{L}{\rm d}\lambda \, h^{N(2)}_{ij}(x^i)\nonumber\\&=a_{\rm rf}\,
    \frac{1}{2}n^in^j\int_0^{L}n_k{\rm d}x^k \, h^{N(2)}_{ij}(x^i)\,,
\end{align}
where we used that since $h^{N(2)}_{ij}(x^i)$ does not depend on time $\int d\lambda =\int  n_k{\rm d}x^k$. The resulting effect is that our sphere of detectors is now displaced by $\Delta L_{\rm i}(x^i) = \frac{1}{2}\int_0^{L}n_k{\rm d}x^k n^in^jh^{N(2)}_{ij}(x^i)$; see Eq.~\eqref{eq:shiftdiscussion} and the surrounding discussion. If we calibrate the detector after generation of the constant strain, we eliminate the memory effect. However, if our geodesic clock were calibrated before strain generation, we should be able to observe the strain buildup, as in the GW memory effect. Moreover, since the constant strain produces a non-trivial Riemann tensor, it may have non-local effects. Thus, despite Ref.~\cite{Domenech:2020xin} showing that the constant mode can be locally gauged away, it is conceivable that it could be observed by extended measurements.

\textit{Conclusion.}---In this work, we computed the time delay caused by passing GWs at second order in cosmological perturbation theory as measured by geodesic observers exchanging EM signals, without gauge fixing. We showed that the time delay measured by a GW detector
is gauge invariant. Then, we identified the effect of passing GWs as that with a quadrupolar dependence on the direction of propagation of the EM signal. We found that, after removing contributions associated with the motion of the observers, the resulting physical GW effect on the time delay at second order is given by the gauge invariant combination  $h_{ij}^{N(2)}$ \eqref{eq:hnij}, which coincides with the TT components of the spatial metric evaluated in the Newton gauge. Therefore, the gauge issue of the induced GW strain is resolved. 

Our result also applies to a more general setting, as we have not made any assumptions about the GW source. Namely, we have demonstrated that GW strain predictions from any cosmic source, performed in the Newton gauge, are free from gauge artifacts and directly related to observables.

Note that, although we assumed geodesic observers, which is an appropriate approximation for PTAs and space-borne interferometers like LISA, our calculations can be generalized to non-geodesic observers like LIGO, by considering the acceleration caused by external forces. However, external forces affect only the subtraction associated with the observer’s motion, not the effect due to propagating GWs.

We have only focused on scalar squared contributions to the second-order transverse-traceless components, since they are the relevant quantities for the (scalar-scalar) induced GWs. But in general, there are also scalar-tensor contributions. In fact, it has been found that scalar-tensor induced GWs suffer from divergences in some cases \cite{Chang:2022vlv,Yu:2023lmo,Bari:2023rcw,Picard:2023sbz,Picard:2025bwq}. If it is a gauge issue, it would be clear in a generalization of the time delay calculation to include scalar-tensor terms. It would also be interesting to study the geodesic deviation equation in terms of local tetrads and confirm that $h_{ij}^{N(2)}$ \eqref{eq:hnij} is the quantity describing the effect of GWs at second order.

We conclude by clarifying that, although we demonstrated that the GW strain measured by a GW detector coincides with one in the Newton gauge, our result does not imply that the energy density associated with GWs at second order coincides with the one computed in the Newton gauge. Namely, there is an additional gauge ambiguity in the definition of the (pseudo) energy momentum tensor of GWs  \cite{Isaacson:1968zza} at second order in cosmological perturbation theory (see, e.g., Ref.~\cite{Cai:2021ndu,Ota:2021fdv} for a study in a fixed time-slice). This, however, is a different problem. It concerns the backreaction of second-order perturbations on the background spacetime, which is important for estimating the expansion rate of the Universe, but is unrelated to the strain measured by a GW detector. We leave this issue for future work.

\textit{Acknowledgements.}---We thank Domenico Giulini, Misao Sasaki, and  Cheng-jun Fang for valuable discussions. 
This work is supported by the National Key Research and Development Program of China, Grant No. 2021YFC2203004, and by the DFG under the Emmy-Noether program project number 496592360. 
G.D. and S.P. also acknowledge support by JSPS KAKENHI grant No. JP24K00624. S.P. and A.W. are supported in part by the National Natural Science Foundation of China (NSFC) Grants Nos. 12475066 and 12447101. A.W. is also supported by the UCAS Joint PhD Training Program.

\bibliography{ref}

\begin{thebibliography}{103}%
\makeatletter
\providecommand \@ifxundefined [1]{%
 \@ifx{#1\undefined}
}%
\providecommand \@ifnum [1]{%
 \ifnum #1\expandafter \@firstoftwo
 \else \expandafter \@secondoftwo
 \fi
}%
\providecommand \@ifx [1]{%
 \ifx #1\expandafter \@firstoftwo
 \else \expandafter \@secondoftwo
 \fi
}%
\providecommand \natexlab [1]{#1}%
\providecommand \enquote  [1]{``#1''}%
\providecommand \bibnamefont  [1]{#1}%
\providecommand \bibfnamefont [1]{#1}%
\providecommand \citenamefont [1]{#1}%
\providecommand \href@noop [0]{\@secondoftwo}%
\providecommand \href [0]{\begingroup \@sanitize@url \@href}%
\providecommand \@href[1]{\@@startlink{#1}\@@href}%
\providecommand \@@href[1]{\endgroup#1\@@endlink}%
\providecommand \@sanitize@url [0]{\catcode `\\12\catcode `\$12\catcode
  `\&12\catcode `\#12\catcode `\^12\catcode `\_12\catcode `\%12\relax}%
\providecommand \@@startlink[1]{}%
\providecommand \@@endlink[0]{}%
\providecommand \url  [0]{\begingroup\@sanitize@url \@url }%
\providecommand \@url [1]{\endgroup\@href {#1}{\urlprefix }}%
\providecommand \urlprefix  [0]{URL }%
\providecommand \Eprint [0]{\href }%
\providecommand \doibase [0]{https://doi.org/}%
\providecommand \selectlanguage [0]{\@gobble}%
\providecommand \bibinfo  [0]{\@secondoftwo}%
\providecommand \bibfield  [0]{\@secondoftwo}%
\providecommand \translation [1]{[#1]}%
\providecommand \BibitemOpen [0]{}%
\providecommand \bibitemStop [0]{}%
\providecommand \bibitemNoStop [0]{.\EOS\space}%
\providecommand \EOS [0]{\spacefactor3000\relax}%
\providecommand \BibitemShut  [1]{\csname bibitem#1\endcsname}%
\let\auto@bib@innerbib\@empty
%</preamble>
\bibitem [{\citenamefont {Abbott}\ \emph {et~al.}(2016)\citenamefont {Abbott}
  \emph {et~al.}}]{LIGOScientific:2016aoc}%
  \BibitemOpen
  \bibfield  {author} {\bibinfo {author} {\bibfnamefont {B.~P.}\ \bibnamefont
  {Abbott}} \emph {et~al.} (\bibinfo {collaboration} {LIGO Scientific,
  Virgo}),\ }\href {https://doi.org/10.1103/PhysRevLett.116.061102} {\bibfield
  {journal} {\bibinfo  {journal} {Phys. Rev. Lett.}\ }\textbf {\bibinfo
  {volume} {116}},\ \bibinfo {pages} {061102} (\bibinfo {year} {2016})},\
  \Eprint {https://arxiv.org/abs/1602.03837} {arXiv:1602.03837 [gr-qc]}
  \BibitemShut {NoStop}%
\bibitem [{\citenamefont {Agazie}\ \emph
  {et~al.}(2023{\natexlab{a}})\citenamefont {Agazie} \emph
  {et~al.}}]{NANOGrav:2023gor}%
  \BibitemOpen
  \bibfield  {author} {\bibinfo {author} {\bibfnamefont {G.}~\bibnamefont
  {Agazie}} \emph {et~al.} (\bibinfo {collaboration} {NANOGrav}),\ }\href
  {https://doi.org/10.3847/2041-8213/acdac6} {\bibfield  {journal} {\bibinfo
  {journal} {Astrophys. J. Lett.}\ }\textbf {\bibinfo {volume} {951}},\
  \bibinfo {pages} {L8} (\bibinfo {year} {2023}{\natexlab{a}})},\ \Eprint
  {https://arxiv.org/abs/2306.16213} {arXiv:2306.16213 [astro-ph.HE]}
  \BibitemShut {NoStop}%
\bibitem [{\citenamefont {Agazie}\ \emph
  {et~al.}(2023{\natexlab{b}})\citenamefont {Agazie} \emph
  {et~al.}}]{NANOGrav:2023hde}%
  \BibitemOpen
  \bibfield  {author} {\bibinfo {author} {\bibfnamefont {G.}~\bibnamefont
  {Agazie}} \emph {et~al.} (\bibinfo {collaboration} {NANOGrav}),\ }\href
  {https://doi.org/10.3847/2041-8213/acda9a} {\bibfield  {journal} {\bibinfo
  {journal} {Astrophys. J. Lett.}\ }\textbf {\bibinfo {volume} {951}},\
  \bibinfo {pages} {L9} (\bibinfo {year} {2023}{\natexlab{b}})},\ \Eprint
  {https://arxiv.org/abs/2306.16217} {arXiv:2306.16217 [astro-ph.HE]}
  \BibitemShut {NoStop}%
\bibitem [{\citenamefont {Antoniadis}\ \emph
  {et~al.}(2023{\natexlab{a}})\citenamefont {Antoniadis} \emph
  {et~al.}}]{EPTA:2023sfo}%
  \BibitemOpen
  \bibfield  {author} {\bibinfo {author} {\bibfnamefont {J.}~\bibnamefont
  {Antoniadis}} \emph {et~al.} (\bibinfo {collaboration} {EPTA}),\ }\href
  {https://doi.org/10.1051/0004-6361/202346841} {\bibfield  {journal} {\bibinfo
   {journal} {Astron. Astrophys.}\ }\textbf {\bibinfo {volume} {678}},\
  \bibinfo {pages} {A48} (\bibinfo {year} {2023}{\natexlab{a}})},\ \Eprint
  {https://arxiv.org/abs/2306.16224} {arXiv:2306.16224 [astro-ph.HE]}
  \BibitemShut {NoStop}%
\bibitem [{\citenamefont {Antoniadis}\ \emph
  {et~al.}(2023{\natexlab{b}})\citenamefont {Antoniadis} \emph
  {et~al.}}]{EPTA:2023fyk}%
  \BibitemOpen
  \bibfield  {author} {\bibinfo {author} {\bibfnamefont {J.}~\bibnamefont
  {Antoniadis}} \emph {et~al.} (\bibinfo {collaboration} {EPTA, InPTA:}),\
  }\href {https://doi.org/10.1051/0004-6361/202346844} {\bibfield  {journal}
  {\bibinfo  {journal} {Astron. Astrophys.}\ }\textbf {\bibinfo {volume}
  {678}},\ \bibinfo {pages} {A50} (\bibinfo {year} {2023}{\natexlab{b}})},\
  \Eprint {https://arxiv.org/abs/2306.16214} {arXiv:2306.16214 [astro-ph.HE]}
  \BibitemShut {NoStop}%
\bibitem [{\citenamefont {Antoniadis}\ \emph {et~al.}(2024)\citenamefont
  {Antoniadis} \emph {et~al.}}]{EPTA:2023xxk}%
  \BibitemOpen
  \bibfield  {author} {\bibinfo {author} {\bibfnamefont {J.}~\bibnamefont
  {Antoniadis}} \emph {et~al.} (\bibinfo {collaboration} {EPTA, InPTA}),\
  }\href {https://doi.org/10.1051/0004-6361/202347433} {\bibfield  {journal}
  {\bibinfo  {journal} {Astron. Astrophys.}\ }\textbf {\bibinfo {volume}
  {685}},\ \bibinfo {pages} {A94} (\bibinfo {year} {2024})},\ \Eprint
  {https://arxiv.org/abs/2306.16227} {arXiv:2306.16227 [astro-ph.CO]}
  \BibitemShut {NoStop}%
\bibitem [{\citenamefont {Reardon}\ \emph
  {et~al.}(2023{\natexlab{a}})\citenamefont {Reardon} \emph
  {et~al.}}]{Reardon:2023gzh}%
  \BibitemOpen
  \bibfield  {author} {\bibinfo {author} {\bibfnamefont {D.~J.}\ \bibnamefont
  {Reardon}} \emph {et~al.},\ }\href {https://doi.org/10.3847/2041-8213/acdd02}
  {\bibfield  {journal} {\bibinfo  {journal} {Astrophys. J. Lett.}\ }\textbf
  {\bibinfo {volume} {951}},\ \bibinfo {pages} {L6} (\bibinfo {year}
  {2023}{\natexlab{a}})},\ \Eprint {https://arxiv.org/abs/2306.16215}
  {arXiv:2306.16215 [astro-ph.HE]} \BibitemShut {NoStop}%
\bibitem [{\citenamefont {Zic}\ \emph {et~al.}(2023)\citenamefont {Zic} \emph
  {et~al.}}]{Zic:2023gta}%
  \BibitemOpen
  \bibfield  {author} {\bibinfo {author} {\bibfnamefont {A.}~\bibnamefont
  {Zic}} \emph {et~al.},\ }\href {https://doi.org/10.1017/pasa.2023.36}
  {\bibfield  {journal} {\bibinfo  {journal} {Publ. Astron. Soc. Austral.}\
  }\textbf {\bibinfo {volume} {40}},\ \bibinfo {pages} {e049} (\bibinfo {year}
  {2023})},\ \Eprint {https://arxiv.org/abs/2306.16230} {arXiv:2306.16230
  [astro-ph.HE]} \BibitemShut {NoStop}%
\bibitem [{\citenamefont {Reardon}\ \emph
  {et~al.}(2023{\natexlab{b}})\citenamefont {Reardon} \emph
  {et~al.}}]{Reardon:2023zen}%
  \BibitemOpen
  \bibfield  {author} {\bibinfo {author} {\bibfnamefont {D.~J.}\ \bibnamefont
  {Reardon}} \emph {et~al.},\ }\href {https://doi.org/10.3847/2041-8213/acdd03}
  {\bibfield  {journal} {\bibinfo  {journal} {Astrophys. J. Lett.}\ }\textbf
  {\bibinfo {volume} {951}},\ \bibinfo {pages} {L7} (\bibinfo {year}
  {2023}{\natexlab{b}})},\ \Eprint {https://arxiv.org/abs/2306.16229}
  {arXiv:2306.16229 [astro-ph.HE]} \BibitemShut {NoStop}%
\bibitem [{\citenamefont {Xu}\ \emph {et~al.}(2023)\citenamefont {Xu} \emph
  {et~al.}}]{Xu:2023wog}%
  \BibitemOpen
  \bibfield  {author} {\bibinfo {author} {\bibfnamefont {H.}~\bibnamefont {Xu}}
  \emph {et~al.},\ }\href {https://doi.org/10.1088/1674-4527/acdfa5} {\bibfield
   {journal} {\bibinfo  {journal} {Res. Astron. Astrophys.}\ }\textbf {\bibinfo
  {volume} {23}},\ \bibinfo {pages} {075024} (\bibinfo {year} {2023})},\
  \Eprint {https://arxiv.org/abs/2306.16216} {arXiv:2306.16216 [astro-ph.HE]}
  \BibitemShut {NoStop}%
\bibitem [{\citenamefont {Binetruy}\ \emph {et~al.}(2012)\citenamefont
  {Binetruy}, \citenamefont {Bohe}, \citenamefont {Caprini},\ and\
  \citenamefont {Dufaux}}]{Binetruy:2012ze}%
  \BibitemOpen
  \bibfield  {author} {\bibinfo {author} {\bibfnamefont {P.}~\bibnamefont
  {Binetruy}}, \bibinfo {author} {\bibfnamefont {A.}~\bibnamefont {Bohe}},
  \bibinfo {author} {\bibfnamefont {C.}~\bibnamefont {Caprini}},\ and\ \bibinfo
  {author} {\bibfnamefont {J.-F.}\ \bibnamefont {Dufaux}},\ }\href
  {https://doi.org/10.1088/1475-7516/2012/06/027} {\bibfield  {journal}
  {\bibinfo  {journal} {JCAP}\ }\textbf {\bibinfo {volume} {06}},\ \bibinfo
  {pages} {027}},\ \Eprint {https://arxiv.org/abs/1201.0983} {arXiv:1201.0983
  [gr-qc]} \BibitemShut {NoStop}%
\bibitem [{\citenamefont {Caprini}\ and\ \citenamefont
  {Figueroa}(2018)}]{Caprini:2018mtu}%
  \BibitemOpen
  \bibfield  {author} {\bibinfo {author} {\bibfnamefont {C.}~\bibnamefont
  {Caprini}}\ and\ \bibinfo {author} {\bibfnamefont {D.~G.}\ \bibnamefont
  {Figueroa}},\ }\href {https://doi.org/10.1088/1361-6382/aac608} {\bibfield
  {journal} {\bibinfo  {journal} {Class. Quant. Grav.}\ }\textbf {\bibinfo
  {volume} {35}},\ \bibinfo {pages} {163001} (\bibinfo {year} {2018})},\
  \Eprint {https://arxiv.org/abs/1801.04268} {arXiv:1801.04268 [astro-ph.CO]}
  \BibitemShut {NoStop}%
\bibitem [{\citenamefont {Dom{\`e}nech}\ and\ \citenamefont
  {Pi}(2022)}]{Domenech:2020ers}%
  \BibitemOpen
  \bibfield  {author} {\bibinfo {author} {\bibfnamefont {G.}~\bibnamefont
  {Dom{\`e}nech}}\ and\ \bibinfo {author} {\bibfnamefont {S.}~\bibnamefont
  {Pi}},\ }\href {https://doi.org/10.1007/s11433-021-1839-6} {\bibfield
  {journal} {\bibinfo  {journal} {Sci. China Phys. Mech. Astron.}\ }\textbf
  {\bibinfo {volume} {65}},\ \bibinfo {pages} {230411} (\bibinfo {year}
  {2022})},\ \Eprint {https://arxiv.org/abs/2010.03976} {arXiv:2010.03976
  [astro-ph.CO]} \BibitemShut {NoStop}%
\bibitem [{\citenamefont {Roshan}\ and\ \citenamefont
  {White}(2025)}]{Roshan:2024qnv}%
  \BibitemOpen
  \bibfield  {author} {\bibinfo {author} {\bibfnamefont {R.}~\bibnamefont
  {Roshan}}\ and\ \bibinfo {author} {\bibfnamefont {G.}~\bibnamefont {White}},\
  }\href {https://doi.org/10.1103/RevModPhys.97.015001} {\bibfield  {journal}
  {\bibinfo  {journal} {Rev. Mod. Phys.}\ }\textbf {\bibinfo {volume} {97}},\
  \bibinfo {pages} {015001} (\bibinfo {year} {2025})},\ \Eprint
  {https://arxiv.org/abs/2401.04388} {arXiv:2401.04388 [hep-ph]} \BibitemShut
  {NoStop}%
\bibitem [{\citenamefont {Dom{\`e}nech}\ \emph {et~al.}(2024)\citenamefont
  {Dom{\`e}nech}, \citenamefont {Pi}, \citenamefont {Wang},\ and\ \citenamefont
  {Wang}}]{Domenech:2024rks}%
  \BibitemOpen
  \bibfield  {author} {\bibinfo {author} {\bibfnamefont {G.}~\bibnamefont
  {Dom{\`e}nech}}, \bibinfo {author} {\bibfnamefont {S.}~\bibnamefont {Pi}},
  \bibinfo {author} {\bibfnamefont {A.}~\bibnamefont {Wang}},\ and\ \bibinfo
  {author} {\bibfnamefont {J.}~\bibnamefont {Wang}},\ }\href
  {https://doi.org/10.1088/1475-7516/2024/08/054} {\bibfield  {journal}
  {\bibinfo  {journal} {JCAP}\ }\textbf {\bibinfo {volume} {08}},\ \bibinfo
  {pages} {054}},\ \Eprint {https://arxiv.org/abs/2402.18965} {arXiv:2402.18965
  [astro-ph.CO]} \BibitemShut {NoStop}%
\bibitem [{\citenamefont {Bian}\ \emph {et~al.}(2025)\citenamefont {Bian} \emph
  {et~al.}}]{Bian:2025ifp}%
  \BibitemOpen
  \bibfield  {author} {\bibinfo {author} {\bibfnamefont {L.}~\bibnamefont
  {Bian}} \emph {et~al.},\ }\href@noop {} {\  (\bibinfo {year} {2025})},\
  \Eprint {https://arxiv.org/abs/2505.19747} {arXiv:2505.19747 [gr-qc]}
  \BibitemShut {NoStop}%
\bibitem [{\citenamefont {Tomita}(1967)}]{Tomita:1967wkp}%
  \BibitemOpen
  \bibfield  {author} {\bibinfo {author} {\bibfnamefont {K.}~\bibnamefont
  {Tomita}},\ }\href {https://doi.org/10.1143/PTP.37.831} {\bibfield  {journal}
  {\bibinfo  {journal} {Prog. Theor. Phys.}\ }\textbf {\bibinfo {volume}
  {37}},\ \bibinfo {pages} {831} (\bibinfo {year} {1967})}\BibitemShut
  {NoStop}%
\bibitem [{\citenamefont {Matarrese}\ \emph {et~al.}(1993)\citenamefont
  {Matarrese}, \citenamefont {Pantano},\ and\ \citenamefont
  {Saez}}]{Matarrese:1992rp}%
  \BibitemOpen
  \bibfield  {author} {\bibinfo {author} {\bibfnamefont {S.}~\bibnamefont
  {Matarrese}}, \bibinfo {author} {\bibfnamefont {O.}~\bibnamefont {Pantano}},\
  and\ \bibinfo {author} {\bibfnamefont {D.}~\bibnamefont {Saez}},\ }\href
  {https://doi.org/10.1103/PhysRevD.47.1311} {\bibfield  {journal} {\bibinfo
  {journal} {Phys. Rev. D}\ }\textbf {\bibinfo {volume} {47}},\ \bibinfo
  {pages} {1311} (\bibinfo {year} {1993})}\BibitemShut {NoStop}%
\bibitem [{\citenamefont {Matarrese}\ \emph {et~al.}(1994)\citenamefont
  {Matarrese}, \citenamefont {Pantano},\ and\ \citenamefont
  {Saez}}]{Matarrese:1993zf}%
  \BibitemOpen
  \bibfield  {author} {\bibinfo {author} {\bibfnamefont {S.}~\bibnamefont
  {Matarrese}}, \bibinfo {author} {\bibfnamefont {O.}~\bibnamefont {Pantano}},\
  and\ \bibinfo {author} {\bibfnamefont {D.}~\bibnamefont {Saez}},\ }\href
  {https://doi.org/10.1103/PhysRevLett.72.320} {\bibfield  {journal} {\bibinfo
  {journal} {Phys. Rev. Lett.}\ }\textbf {\bibinfo {volume} {72}},\ \bibinfo
  {pages} {320} (\bibinfo {year} {1994})},\ \Eprint
  {https://arxiv.org/abs/astro-ph/9310036} {arXiv:astro-ph/9310036}
  \BibitemShut {NoStop}%
\bibitem [{\citenamefont {Bruni}\ \emph {et~al.}(1997)\citenamefont {Bruni},
  \citenamefont {Matarrese}, \citenamefont {Mollerach},\ and\ \citenamefont
  {Sonego}}]{Bruni:1996im}%
  \BibitemOpen
  \bibfield  {author} {\bibinfo {author} {\bibfnamefont {M.}~\bibnamefont
  {Bruni}}, \bibinfo {author} {\bibfnamefont {S.}~\bibnamefont {Matarrese}},
  \bibinfo {author} {\bibfnamefont {S.}~\bibnamefont {Mollerach}},\ and\
  \bibinfo {author} {\bibfnamefont {S.}~\bibnamefont {Sonego}},\ }\href
  {https://doi.org/10.1088/0264-9381/14/9/014} {\bibfield  {journal} {\bibinfo
  {journal} {Class. Quant. Grav.}\ }\textbf {\bibinfo {volume} {14}},\ \bibinfo
  {pages} {2585} (\bibinfo {year} {1997})},\ \Eprint
  {https://arxiv.org/abs/gr-qc/9609040} {arXiv:gr-qc/9609040} \BibitemShut
  {NoStop}%
\bibitem [{\citenamefont {Matarrese}\ and\ \citenamefont
  {Mollerach}(1996)}]{Matarrese:1996pp}%
  \BibitemOpen
  \bibfield  {author} {\bibinfo {author} {\bibfnamefont {S.}~\bibnamefont
  {Matarrese}}\ and\ \bibinfo {author} {\bibfnamefont {S.}~\bibnamefont
  {Mollerach}},\ }in\ \href@noop {} {\emph {\bibinfo {booktitle} {{ERE -
  Spanish Relativity Conference}}}}\ (\bibinfo {year} {1996})\ \Eprint
  {https://arxiv.org/abs/astro-ph/9705168} {arXiv:astro-ph/9705168}
  \BibitemShut {NoStop}%
\bibitem [{\citenamefont {Matarrese}\ \emph {et~al.}(1998)\citenamefont
  {Matarrese}, \citenamefont {Mollerach},\ and\ \citenamefont
  {Bruni}}]{Matarrese:1997ay}%
  \BibitemOpen
  \bibfield  {author} {\bibinfo {author} {\bibfnamefont {S.}~\bibnamefont
  {Matarrese}}, \bibinfo {author} {\bibfnamefont {S.}~\bibnamefont
  {Mollerach}},\ and\ \bibinfo {author} {\bibfnamefont {M.}~\bibnamefont
  {Bruni}},\ }\href {https://doi.org/10.1103/PhysRevD.58.043504} {\bibfield
  {journal} {\bibinfo  {journal} {Phys. Rev. D}\ }\textbf {\bibinfo {volume}
  {58}},\ \bibinfo {pages} {043504} (\bibinfo {year} {1998})},\ \Eprint
  {https://arxiv.org/abs/astro-ph/9707278} {arXiv:astro-ph/9707278}
  \BibitemShut {NoStop}%
\bibitem [{\citenamefont {Ananda}\ \emph {et~al.}(2007)\citenamefont {Ananda},
  \citenamefont {Clarkson},\ and\ \citenamefont {Wands}}]{Ananda:2006af}%
  \BibitemOpen
  \bibfield  {author} {\bibinfo {author} {\bibfnamefont {K.~N.}\ \bibnamefont
  {Ananda}}, \bibinfo {author} {\bibfnamefont {C.}~\bibnamefont {Clarkson}},\
  and\ \bibinfo {author} {\bibfnamefont {D.}~\bibnamefont {Wands}},\ }\href
  {https://doi.org/10.1103/PhysRevD.75.123518} {\bibfield  {journal} {\bibinfo
  {journal} {Phys. Rev. D}\ }\textbf {\bibinfo {volume} {75}},\ \bibinfo
  {pages} {123518} (\bibinfo {year} {2007})},\ \Eprint
  {https://arxiv.org/abs/gr-qc/0612013} {arXiv:gr-qc/0612013} \BibitemShut
  {NoStop}%
\bibitem [{\citenamefont {Baumann}\ \emph {et~al.}(2007)\citenamefont
  {Baumann}, \citenamefont {Steinhardt}, \citenamefont {Takahashi},\ and\
  \citenamefont {Ichiki}}]{Baumann:2007zm}%
  \BibitemOpen
  \bibfield  {author} {\bibinfo {author} {\bibfnamefont {D.}~\bibnamefont
  {Baumann}}, \bibinfo {author} {\bibfnamefont {P.~J.}\ \bibnamefont
  {Steinhardt}}, \bibinfo {author} {\bibfnamefont {K.}~\bibnamefont
  {Takahashi}},\ and\ \bibinfo {author} {\bibfnamefont {K.}~\bibnamefont
  {Ichiki}},\ }\href {https://doi.org/10.1103/PhysRevD.76.084019} {\bibfield
  {journal} {\bibinfo  {journal} {Phys. Rev. D}\ }\textbf {\bibinfo {volume}
  {76}},\ \bibinfo {pages} {084019} (\bibinfo {year} {2007})},\ \Eprint
  {https://arxiv.org/abs/hep-th/0703290} {arXiv:hep-th/0703290} \BibitemShut
  {NoStop}%
\bibitem [{\citenamefont {Yuan}\ and\ \citenamefont
  {Huang}(2021)}]{Yuan:2021qgz}%
  \BibitemOpen
  \bibfield  {author} {\bibinfo {author} {\bibfnamefont {C.}~\bibnamefont
  {Yuan}}\ and\ \bibinfo {author} {\bibfnamefont {Q.-G.}\ \bibnamefont
  {Huang}},\ }\href {https://doi.org/10.1016/j.isci.2021.102860} {\bibfield
  {journal} {\bibinfo  {journal} {iScience}\ }\textbf {\bibinfo {volume}
  {24}},\ \bibinfo {pages} {102860} (\bibinfo {year} {2021})},\ \Eprint
  {https://arxiv.org/abs/2103.04739} {arXiv:2103.04739 [astro-ph.GA]}
  \BibitemShut {NoStop}%
\bibitem [{\citenamefont {Dom{\`e}nech}(2021)}]{Domenech:2021ztg}%
  \BibitemOpen
  \bibfield  {author} {\bibinfo {author} {\bibfnamefont {G.}~\bibnamefont
  {Dom{\`e}nech}},\ }\href {https://doi.org/10.3390/universe7110398} {\bibfield
   {journal} {\bibinfo  {journal} {Universe}\ }\textbf {\bibinfo {volume}
  {7}},\ \bibinfo {pages} {398} (\bibinfo {year} {2021})},\ \Eprint
  {https://arxiv.org/abs/2109.01398} {arXiv:2109.01398 [gr-qc]} \BibitemShut
  {NoStop}%
\bibitem [{\citenamefont {Aghanim}\ \emph {et~al.}(2020)\citenamefont {Aghanim}
  \emph {et~al.}}]{Planck:2018vyg}%
  \BibitemOpen
  \bibfield  {author} {\bibinfo {author} {\bibfnamefont {N.}~\bibnamefont
  {Aghanim}} \emph {et~al.} (\bibinfo {collaboration} {Planck}),\ }\href
  {https://doi.org/10.1051/0004-6361/201833910} {\bibfield  {journal} {\bibinfo
   {journal} {Astron. Astrophys.}\ }\textbf {\bibinfo {volume} {641}},\
  \bibinfo {pages} {A6} (\bibinfo {year} {2020})},\ \bibinfo {note} {[Erratum:
  Astron.Astrophys. 652, C4 (2021)]},\ \Eprint
  {https://arxiv.org/abs/1807.06209} {arXiv:1807.06209 [astro-ph.CO]}
  \BibitemShut {NoStop}%
\bibitem [{\citenamefont {Saito}\ and\ \citenamefont
  {Yokoyama}(2009)}]{Saito:2008jc}%
  \BibitemOpen
  \bibfield  {author} {\bibinfo {author} {\bibfnamefont {R.}~\bibnamefont
  {Saito}}\ and\ \bibinfo {author} {\bibfnamefont {J.}~\bibnamefont
  {Yokoyama}},\ }\href {https://doi.org/10.1103/PhysRevLett.102.161101}
  {\bibfield  {journal} {\bibinfo  {journal} {Phys. Rev. Lett.}\ }\textbf
  {\bibinfo {volume} {102}},\ \bibinfo {pages} {161101} (\bibinfo {year}
  {2009})},\ \bibinfo {note} {[Erratum: Phys.Rev.Lett. 107, 069901 (2011)]},\
  \Eprint {https://arxiv.org/abs/0812.4339} {arXiv:0812.4339 [astro-ph]}
  \BibitemShut {NoStop}%
\bibitem [{\citenamefont {Cai}\ \emph {et~al.}(2019)\citenamefont {Cai},
  \citenamefont {Pi},\ and\ \citenamefont {Sasaki}}]{Cai:2018dig}%
  \BibitemOpen
  \bibfield  {author} {\bibinfo {author} {\bibfnamefont {R.-g.}\ \bibnamefont
  {Cai}}, \bibinfo {author} {\bibfnamefont {S.}~\bibnamefont {Pi}},\ and\
  \bibinfo {author} {\bibfnamefont {M.}~\bibnamefont {Sasaki}},\ }\href
  {https://doi.org/10.1103/PhysRevLett.122.201101} {\bibfield  {journal}
  {\bibinfo  {journal} {Phys. Rev. Lett.}\ }\textbf {\bibinfo {volume} {122}},\
  \bibinfo {pages} {201101} (\bibinfo {year} {2019})},\ \Eprint
  {https://arxiv.org/abs/1810.11000} {arXiv:1810.11000 [astro-ph.CO]}
  \BibitemShut {NoStop}%
\bibitem [{\citenamefont {Byrnes}\ \emph {et~al.}(2019)\citenamefont {Byrnes},
  \citenamefont {Cole},\ and\ \citenamefont {Patil}}]{Byrnes:2018txb}%
  \BibitemOpen
  \bibfield  {author} {\bibinfo {author} {\bibfnamefont {C.~T.}\ \bibnamefont
  {Byrnes}}, \bibinfo {author} {\bibfnamefont {P.~S.}\ \bibnamefont {Cole}},\
  and\ \bibinfo {author} {\bibfnamefont {S.~P.}\ \bibnamefont {Patil}},\ }\href
  {https://doi.org/10.1088/1475-7516/2019/06/028} {\bibfield  {journal}
  {\bibinfo  {journal} {JCAP}\ }\textbf {\bibinfo {volume} {06}},\ \bibinfo
  {pages} {028}},\ \Eprint {https://arxiv.org/abs/1811.11158} {arXiv:1811.11158
  [astro-ph.CO]} \BibitemShut {NoStop}%
\bibitem [{\citenamefont {Chen}\ \emph {et~al.}(2020)\citenamefont {Chen},
  \citenamefont {Yuan},\ and\ \citenamefont {Huang}}]{Chen:2019xse}%
  \BibitemOpen
  \bibfield  {author} {\bibinfo {author} {\bibfnamefont {Z.-C.}\ \bibnamefont
  {Chen}}, \bibinfo {author} {\bibfnamefont {C.}~\bibnamefont {Yuan}},\ and\
  \bibinfo {author} {\bibfnamefont {Q.-G.}\ \bibnamefont {Huang}},\ }\href
  {https://doi.org/10.1103/PhysRevLett.124.251101} {\bibfield  {journal}
  {\bibinfo  {journal} {Phys. Rev. Lett.}\ }\textbf {\bibinfo {volume} {124}},\
  \bibinfo {pages} {25} (\bibinfo {year} {2020})},\ \Eprint
  {https://arxiv.org/abs/1910.12239} {arXiv:1910.12239 [astro-ph.CO]}
  \BibitemShut {NoStop}%
\bibitem [{\citenamefont {Ren}\ \emph {et~al.}(2023)\citenamefont {Ren},
  \citenamefont {Zhao}, \citenamefont {Cao}, \citenamefont {Guo}, \citenamefont
  {Han}, \citenamefont {Jin},\ and\ \citenamefont {Wu}}]{Ren:2023yec}%
  \BibitemOpen
  \bibfield  {author} {\bibinfo {author} {\bibfnamefont {Z.}~\bibnamefont
  {Ren}}, \bibinfo {author} {\bibfnamefont {T.}~\bibnamefont {Zhao}}, \bibinfo
  {author} {\bibfnamefont {Z.}~\bibnamefont {Cao}}, \bibinfo {author}
  {\bibfnamefont {Z.-K.}\ \bibnamefont {Guo}}, \bibinfo {author} {\bibfnamefont
  {W.-B.}\ \bibnamefont {Han}}, \bibinfo {author} {\bibfnamefont {H.-B.}\
  \bibnamefont {Jin}},\ and\ \bibinfo {author} {\bibfnamefont {Y.-L.}\
  \bibnamefont {Wu}},\ }\href {https://doi.org/10.1007/s11467-023-1318-y}
  {\bibfield  {journal} {\bibinfo  {journal} {Front. Phys. (Beijing)}\ }\textbf
  {\bibinfo {volume} {18}},\ \bibinfo {pages} {64302} (\bibinfo {year}
  {2023})},\ \Eprint {https://arxiv.org/abs/2301.02967} {arXiv:2301.02967
  [gr-qc]} \BibitemShut {NoStop}%
\bibitem [{\citenamefont {Dandoy}\ \emph {et~al.}(2023)\citenamefont {Dandoy},
  \citenamefont {Domcke},\ and\ \citenamefont {Rompineve}}]{Dandoy:2023jot}%
  \BibitemOpen
  \bibfield  {author} {\bibinfo {author} {\bibfnamefont {V.}~\bibnamefont
  {Dandoy}}, \bibinfo {author} {\bibfnamefont {V.}~\bibnamefont {Domcke}},\
  and\ \bibinfo {author} {\bibfnamefont {F.}~\bibnamefont {Rompineve}},\ }\href
  {https://doi.org/10.21468/SciPostPhysCore.6.3.060} {\bibfield  {journal}
  {\bibinfo  {journal} {SciPost Phys. Core}\ }\textbf {\bibinfo {volume} {6}},\
  \bibinfo {pages} {060} (\bibinfo {year} {2023})},\ \Eprint
  {https://arxiv.org/abs/2302.07901} {arXiv:2302.07901 [astro-ph.CO]}
  \BibitemShut {NoStop}%
\bibitem [{\citenamefont {Bagui}\ \emph {et~al.}(2025)\citenamefont {Bagui}
  \emph {et~al.}}]{LISACosmologyWorkingGroup:2023njw}%
  \BibitemOpen
  \bibfield  {author} {\bibinfo {author} {\bibfnamefont {E.}~\bibnamefont
  {Bagui}} \emph {et~al.} (\bibinfo {collaboration} {LISA Cosmology Working
  Group}),\ }\href {https://doi.org/10.1007/s41114-024-00053-w} {\bibfield
  {journal} {\bibinfo  {journal} {Living Rev. Rel.}\ }\textbf {\bibinfo
  {volume} {28}},\ \bibinfo {pages} {1} (\bibinfo {year} {2025})},\ \Eprint
  {https://arxiv.org/abs/2310.19857} {arXiv:2310.19857 [astro-ph.CO]}
  \BibitemShut {NoStop}%
\bibitem [{\citenamefont {{\"O}zsoy}\ and\ \citenamefont
  {Tasinato}(2023)}]{Ozsoy:2023ryl}%
  \BibitemOpen
  \bibfield  {author} {\bibinfo {author} {\bibfnamefont {O.}~\bibnamefont
  {{\"O}zsoy}}\ and\ \bibinfo {author} {\bibfnamefont {G.}~\bibnamefont
  {Tasinato}},\ }\href {https://doi.org/10.3390/universe9050203} {\bibfield
  {journal} {\bibinfo  {journal} {Universe}\ }\textbf {\bibinfo {volume} {9}},\
  \bibinfo {pages} {203} (\bibinfo {year} {2023})},\ \Eprint
  {https://arxiv.org/abs/2301.03600} {arXiv:2301.03600 [astro-ph.CO]}
  \BibitemShut {NoStop}%
\bibitem [{\citenamefont {Iovino}\ \emph {et~al.}(2024)\citenamefont {Iovino},
  \citenamefont {Perna}, \citenamefont {Riotto},\ and\ \citenamefont
  {Veerm{\"a}e}}]{Iovino:2024tyg}%
  \BibitemOpen
  \bibfield  {author} {\bibinfo {author} {\bibfnamefont {A.~J.}\ \bibnamefont
  {Iovino}}, \bibinfo {author} {\bibfnamefont {G.}~\bibnamefont {Perna}},
  \bibinfo {author} {\bibfnamefont {A.}~\bibnamefont {Riotto}},\ and\ \bibinfo
  {author} {\bibfnamefont {H.}~\bibnamefont {Veerm{\"a}e}},\ }\href
  {https://doi.org/10.1088/1475-7516/2024/10/050} {\bibfield  {journal}
  {\bibinfo  {journal} {JCAP}\ }\textbf {\bibinfo {volume} {10}},\ \bibinfo
  {pages} {050}},\ \Eprint {https://arxiv.org/abs/2406.20089} {arXiv:2406.20089
  [astro-ph.CO]} \BibitemShut {NoStop}%
\bibitem [{\citenamefont {Luo}\ \emph {et~al.}(2025)\citenamefont {Luo} \emph
  {et~al.}}]{Luo:2025ewp}%
  \BibitemOpen
  \bibfield  {author} {\bibinfo {author} {\bibfnamefont {J.}~\bibnamefont
  {Luo}} \emph {et~al.},\ }\href@noop {} {\  (\bibinfo {year} {2025})},\
  \Eprint {https://arxiv.org/abs/2502.20138} {arXiv:2502.20138 [gr-qc]}
  \BibitemShut {NoStop}%
\bibitem [{\citenamefont {Iovino}\ \emph {et~al.}(2025)\citenamefont {Iovino},
  \citenamefont {Perna},\ and\ \citenamefont {Veerm{\"a}e}}]{Iovino:2025cdy}%
  \BibitemOpen
  \bibfield  {author} {\bibinfo {author} {\bibfnamefont {A.}~\bibnamefont
  {Iovino}, \bibfnamefont {Junior.}}, \bibinfo {author} {\bibfnamefont
  {G.}~\bibnamefont {Perna}},\ and\ \bibinfo {author} {\bibfnamefont
  {H.}~\bibnamefont {Veerm{\"a}e}},\ }\href@noop {} {\  (\bibinfo {year}
  {2025})},\ \Eprint {https://arxiv.org/abs/2512.13648} {arXiv:2512.13648
  [astro-ph.CO]} \BibitemShut {NoStop}%
\bibitem [{\citenamefont {Cai}\ \emph {et~al.}(2020)\citenamefont {Cai},
  \citenamefont {Pi},\ and\ \citenamefont {Sasaki}}]{Cai:2019cdl}%
  \BibitemOpen
  \bibfield  {author} {\bibinfo {author} {\bibfnamefont {R.-G.}\ \bibnamefont
  {Cai}}, \bibinfo {author} {\bibfnamefont {S.}~\bibnamefont {Pi}},\ and\
  \bibinfo {author} {\bibfnamefont {M.}~\bibnamefont {Sasaki}},\ }\href
  {https://doi.org/10.1103/PhysRevD.102.083528} {\bibfield  {journal} {\bibinfo
   {journal} {Phys. Rev. D}\ }\textbf {\bibinfo {volume} {102}},\ \bibinfo
  {pages} {083528} (\bibinfo {year} {2020})},\ \Eprint
  {https://arxiv.org/abs/1909.13728} {arXiv:1909.13728 [astro-ph.CO]}
  \BibitemShut {NoStop}%
\bibitem [{\citenamefont {Dom{\`e}nech}\ \emph {et~al.}(2020)\citenamefont
  {Dom{\`e}nech}, \citenamefont {Pi},\ and\ \citenamefont
  {Sasaki}}]{Domenech:2020kqm}%
  \BibitemOpen
  \bibfield  {author} {\bibinfo {author} {\bibfnamefont {G.}~\bibnamefont
  {Dom{\`e}nech}}, \bibinfo {author} {\bibfnamefont {S.}~\bibnamefont {Pi}},\
  and\ \bibinfo {author} {\bibfnamefont {M.}~\bibnamefont {Sasaki}},\ }\href
  {https://doi.org/10.1088/1475-7516/2020/08/017} {\bibfield  {journal}
  {\bibinfo  {journal} {JCAP}\ }\textbf {\bibinfo {volume} {08}},\ \bibinfo
  {pages} {017}},\ \Eprint {https://arxiv.org/abs/2005.12314} {arXiv:2005.12314
  [gr-qc]} \BibitemShut {NoStop}%
\bibitem [{\citenamefont {Dom{\`e}nech}\ \emph {et~al.}(2021)\citenamefont
  {Dom{\`e}nech}, \citenamefont {Lin},\ and\ \citenamefont
  {Sasaki}}]{Domenech:2020ssp}%
  \BibitemOpen
  \bibfield  {author} {\bibinfo {author} {\bibfnamefont {G.}~\bibnamefont
  {Dom{\`e}nech}}, \bibinfo {author} {\bibfnamefont {C.}~\bibnamefont {Lin}},\
  and\ \bibinfo {author} {\bibfnamefont {M.}~\bibnamefont {Sasaki}},\ }\href
  {https://doi.org/10.1088/1475-7516/2021/11/E01} {\bibfield  {journal}
  {\bibinfo  {journal} {JCAP}\ }\textbf {\bibinfo {volume} {04}},\ \bibinfo
  {pages} {062}},\ \bibinfo {note} {[Erratum: JCAP 11, E01 (2021)]},\ \Eprint
  {https://arxiv.org/abs/2012.08151} {arXiv:2012.08151 [gr-qc]} \BibitemShut
  {NoStop}%
\bibitem [{\citenamefont {Liu}\ \emph {et~al.}(2023)\citenamefont {Liu},
  \citenamefont {Chen},\ and\ \citenamefont {Huang}}]{Liu:2023pau}%
  \BibitemOpen
  \bibfield  {author} {\bibinfo {author} {\bibfnamefont {L.}~\bibnamefont
  {Liu}}, \bibinfo {author} {\bibfnamefont {Z.-C.}\ \bibnamefont {Chen}},\ and\
  \bibinfo {author} {\bibfnamefont {Q.-G.}\ \bibnamefont {Huang}},\ }\href
  {https://doi.org/10.1088/1475-7516/2023/11/071} {\bibfield  {journal}
  {\bibinfo  {journal} {JCAP}\ }\textbf {\bibinfo {volume} {11}},\ \bibinfo
  {pages} {071}},\ \Eprint {https://arxiv.org/abs/2307.14911} {arXiv:2307.14911
  [astro-ph.CO]} \BibitemShut {NoStop}%
\bibitem [{\citenamefont {Hwang}\ \emph {et~al.}(2017)\citenamefont {Hwang},
  \citenamefont {Jeong},\ and\ \citenamefont {Noh}}]{Hwang:2017oxa}%
  \BibitemOpen
  \bibfield  {author} {\bibinfo {author} {\bibfnamefont {J.-C.}\ \bibnamefont
  {Hwang}}, \bibinfo {author} {\bibfnamefont {D.}~\bibnamefont {Jeong}},\ and\
  \bibinfo {author} {\bibfnamefont {H.}~\bibnamefont {Noh}},\ }\href
  {https://doi.org/10.3847/1538-4357/aa74be} {\bibfield  {journal} {\bibinfo
  {journal} {Astrophys. J.}\ }\textbf {\bibinfo {volume} {842}},\ \bibinfo
  {pages} {46} (\bibinfo {year} {2017})},\ \Eprint
  {https://arxiv.org/abs/1704.03500} {arXiv:1704.03500 [astro-ph.CO]}
  \BibitemShut {NoStop}%
\bibitem [{\citenamefont {Maggiore}(2007)}]{Maggiore:2007ulw}%
  \BibitemOpen
  \bibfield  {author} {\bibinfo {author} {\bibfnamefont {M.}~\bibnamefont
  {Maggiore}},\ }\href
  {https://doi.org/10.1093/acprof:oso/9780198570745.001.0001} {\emph {\bibinfo
  {title} {{Gravitational Waves. Vol. 1: Theory and Experiments}}}}\ (\bibinfo
  {publisher} {Oxford University Press},\ \bibinfo {year} {2007})\BibitemShut
  {NoStop}%
\bibitem [{\citenamefont {Tomikawa}\ and\ \citenamefont
  {Kobayashi}(2020)}]{Tomikawa:2019tvi}%
  \BibitemOpen
  \bibfield  {author} {\bibinfo {author} {\bibfnamefont {K.}~\bibnamefont
  {Tomikawa}}\ and\ \bibinfo {author} {\bibfnamefont {T.}~\bibnamefont
  {Kobayashi}},\ }\href {https://doi.org/10.1103/PhysRevD.101.083529}
  {\bibfield  {journal} {\bibinfo  {journal} {Phys. Rev. D}\ }\textbf {\bibinfo
  {volume} {101}},\ \bibinfo {pages} {083529} (\bibinfo {year} {2020})},\
  \Eprint {https://arxiv.org/abs/1910.01880} {arXiv:1910.01880 [gr-qc]}
  \BibitemShut {NoStop}%
\bibitem [{\citenamefont {Gong}(2022)}]{Gong:2019mui}%
  \BibitemOpen
  \bibfield  {author} {\bibinfo {author} {\bibfnamefont {J.-O.}\ \bibnamefont
  {Gong}},\ }\href {https://doi.org/10.3847/1538-4357/ac3a6c} {\bibfield
  {journal} {\bibinfo  {journal} {Astrophys. J.}\ }\textbf {\bibinfo {volume}
  {925}},\ \bibinfo {pages} {102} (\bibinfo {year} {2022})},\ \Eprint
  {https://arxiv.org/abs/1909.12708} {arXiv:1909.12708 [gr-qc]} \BibitemShut
  {NoStop}%
\bibitem [{\citenamefont {De~Luca}\ \emph {et~al.}(2020)\citenamefont
  {De~Luca}, \citenamefont {Franciolini}, \citenamefont {Kehagias},\ and\
  \citenamefont {Riotto}}]{DeLuca:2019ufz}%
  \BibitemOpen
  \bibfield  {author} {\bibinfo {author} {\bibfnamefont {V.}~\bibnamefont
  {De~Luca}}, \bibinfo {author} {\bibfnamefont {G.}~\bibnamefont
  {Franciolini}}, \bibinfo {author} {\bibfnamefont {A.}~\bibnamefont
  {Kehagias}},\ and\ \bibinfo {author} {\bibfnamefont {A.}~\bibnamefont
  {Riotto}},\ }\href {https://doi.org/10.1088/1475-7516/2020/03/014} {\bibfield
   {journal} {\bibinfo  {journal} {JCAP}\ }\textbf {\bibinfo {volume} {03}},\
  \bibinfo {pages} {014}},\ \Eprint {https://arxiv.org/abs/1911.09689}
  {arXiv:1911.09689 [gr-qc]} \BibitemShut {NoStop}%
\bibitem [{\citenamefont {Inomata}\ and\ \citenamefont
  {Terada}(2020)}]{Inomata:2019yww}%
  \BibitemOpen
  \bibfield  {author} {\bibinfo {author} {\bibfnamefont {K.}~\bibnamefont
  {Inomata}}\ and\ \bibinfo {author} {\bibfnamefont {T.}~\bibnamefont
  {Terada}},\ }\href {https://doi.org/10.1103/PhysRevD.101.023523} {\bibfield
  {journal} {\bibinfo  {journal} {Phys. Rev. D}\ }\textbf {\bibinfo {volume}
  {101}},\ \bibinfo {pages} {023523} (\bibinfo {year} {2020})},\ \Eprint
  {https://arxiv.org/abs/1912.00785} {arXiv:1912.00785 [gr-qc]} \BibitemShut
  {NoStop}%
\bibitem [{\citenamefont {Yuan}\ \emph {et~al.}(2020)\citenamefont {Yuan},
  \citenamefont {Chen},\ and\ \citenamefont {Huang}}]{Yuan:2019fwv}%
  \BibitemOpen
  \bibfield  {author} {\bibinfo {author} {\bibfnamefont {C.}~\bibnamefont
  {Yuan}}, \bibinfo {author} {\bibfnamefont {Z.-C.}\ \bibnamefont {Chen}},\
  and\ \bibinfo {author} {\bibfnamefont {Q.-G.}\ \bibnamefont {Huang}},\ }\href
  {https://doi.org/10.1103/PhysRevD.101.063018} {\bibfield  {journal} {\bibinfo
   {journal} {Phys. Rev. D}\ }\textbf {\bibinfo {volume} {101}},\ \bibinfo
  {pages} {6} (\bibinfo {year} {2020})},\ \Eprint
  {https://arxiv.org/abs/1912.00885} {arXiv:1912.00885 [astro-ph.CO]}
  \BibitemShut {NoStop}%
\bibitem [{\citenamefont {Chang}\ \emph {et~al.}(2020)\citenamefont {Chang},
  \citenamefont {Wang},\ and\ \citenamefont {Zhu}}]{Chang:2020iji}%
  \BibitemOpen
  \bibfield  {author} {\bibinfo {author} {\bibfnamefont {Z.}~\bibnamefont
  {Chang}}, \bibinfo {author} {\bibfnamefont {S.}~\bibnamefont {Wang}},\ and\
  \bibinfo {author} {\bibfnamefont {Q.-H.}\ \bibnamefont {Zhu}},\ }\href@noop
  {} {\  (\bibinfo {year} {2020})},\ \Eprint {https://arxiv.org/abs/2009.11994}
  {arXiv:2009.11994 [gr-qc]} \BibitemShut {NoStop}%
\bibitem [{\citenamefont {Dom{\`e}nech}\ and\ \citenamefont
  {Sasaki}(2021)}]{Domenech:2020xin}%
  \BibitemOpen
  \bibfield  {author} {\bibinfo {author} {\bibfnamefont {G.}~\bibnamefont
  {Dom{\`e}nech}}\ and\ \bibinfo {author} {\bibfnamefont {M.}~\bibnamefont
  {Sasaki}},\ }\href {https://doi.org/10.1103/PhysRevD.103.063531} {\bibfield
  {journal} {\bibinfo  {journal} {Phys. Rev. D}\ }\textbf {\bibinfo {volume}
  {103}},\ \bibinfo {pages} {063531} (\bibinfo {year} {2021})},\ \Eprint
  {https://arxiv.org/abs/2012.14016} {arXiv:2012.14016 [gr-qc]} \BibitemShut
  {NoStop}%
\bibitem [{\citenamefont {Ota}\ \emph {et~al.}(2022)\citenamefont {Ota},
  \citenamefont {Macpherson},\ and\ \citenamefont {Coulton}}]{Ota:2021fdv}%
  \BibitemOpen
  \bibfield  {author} {\bibinfo {author} {\bibfnamefont {A.}~\bibnamefont
  {Ota}}, \bibinfo {author} {\bibfnamefont {H.~J.}\ \bibnamefont
  {Macpherson}},\ and\ \bibinfo {author} {\bibfnamefont {W.~R.}\ \bibnamefont
  {Coulton}},\ }\href {https://doi.org/10.1103/PhysRevD.106.063521} {\bibfield
  {journal} {\bibinfo  {journal} {Phys. Rev. D}\ }\textbf {\bibinfo {volume}
  {106}},\ \bibinfo {pages} {063521} (\bibinfo {year} {2022})},\ \Eprint
  {https://arxiv.org/abs/2111.09163} {arXiv:2111.09163 [gr-qc]} \BibitemShut
  {NoStop}%
\bibitem [{\citenamefont {Yuan}\ \emph
  {et~al.}(2025{\natexlab{a}})\citenamefont {Yuan}, \citenamefont {Chen},\ and\
  \citenamefont {Liu}}]{Yuan:2024qfz}%
  \BibitemOpen
  \bibfield  {author} {\bibinfo {author} {\bibfnamefont {C.}~\bibnamefont
  {Yuan}}, \bibinfo {author} {\bibfnamefont {Z.-C.}\ \bibnamefont {Chen}},\
  and\ \bibinfo {author} {\bibfnamefont {L.}~\bibnamefont {Liu}},\ }\href
  {https://doi.org/10.1103/PhysRevD.111.103528} {\bibfield  {journal} {\bibinfo
   {journal} {Phys. Rev. D}\ }\textbf {\bibinfo {volume} {111}},\ \bibinfo
  {pages} {103528} (\bibinfo {year} {2025}{\natexlab{a}})},\ \Eprint
  {https://arxiv.org/abs/2410.18996} {arXiv:2410.18996 [gr-qc]} \BibitemShut
  {NoStop}%
\bibitem [{\citenamefont {Dom{\`e}nech}\ and\ \citenamefont
  {Chluba}(2025)}]{Domenech:2025bvr}%
  \BibitemOpen
  \bibfield  {author} {\bibinfo {author} {\bibfnamefont {G.}~\bibnamefont
  {Dom{\`e}nech}}\ and\ \bibinfo {author} {\bibfnamefont {J.}~\bibnamefont
  {Chluba}},\ }\href {https://doi.org/10.1088/1475-7516/2025/07/034} {\bibfield
   {journal} {\bibinfo  {journal} {JCAP}\ }\textbf {\bibinfo {volume} {07}},\
  \bibinfo {pages} {034}},\ \Eprint {https://arxiv.org/abs/2503.13670}
  {arXiv:2503.13670 [gr-qc]} \BibitemShut {NoStop}%
\bibitem [{\citenamefont {Dom{\`e}nech}\ and\ \citenamefont
  {Sasaki}(2018)}]{Domenech:2017ems}%
  \BibitemOpen
  \bibfield  {author} {\bibinfo {author} {\bibfnamefont {G.}~\bibnamefont
  {Dom{\`e}nech}}\ and\ \bibinfo {author} {\bibfnamefont {M.}~\bibnamefont
  {Sasaki}},\ }\href {https://doi.org/10.1103/PhysRevD.97.023521} {\bibfield
  {journal} {\bibinfo  {journal} {Phys. Rev. D}\ }\textbf {\bibinfo {volume}
  {97}},\ \bibinfo {pages} {023521} (\bibinfo {year} {2018})},\ \Eprint
  {https://arxiv.org/abs/1709.09804} {arXiv:1709.09804 [gr-qc]} \BibitemShut
  {NoStop}%
\bibitem [{\citenamefont {Yuan}\ \emph
  {et~al.}(2025{\natexlab{b}})\citenamefont {Yuan}, \citenamefont {Lu},
  \citenamefont {Chen},\ and\ \citenamefont {Liu}}]{Yuan:2025seu}%
  \BibitemOpen
  \bibfield  {author} {\bibinfo {author} {\bibfnamefont {C.}~\bibnamefont
  {Yuan}}, \bibinfo {author} {\bibfnamefont {Y.}~\bibnamefont {Lu}}, \bibinfo
  {author} {\bibfnamefont {Z.-C.}\ \bibnamefont {Chen}},\ and\ \bibinfo
  {author} {\bibfnamefont {L.}~\bibnamefont {Liu}},\ }\href
  {https://doi.org/10.1088/1475-7516/2025/07/016} {\bibfield  {journal}
  {\bibinfo  {journal} {JCAP}\ }\textbf {\bibinfo {volume} {07}},\ \bibinfo
  {pages} {016}},\ \Eprint {https://arxiv.org/abs/2501.13691} {arXiv:2501.13691
  [astro-ph.CO]} \BibitemShut {NoStop}%
\bibitem [{\citenamefont {Marzke}\ and\ \citenamefont
  {Wheeler}(1964)}]{marzke1964gravitation}%
  \BibitemOpen
  \bibfield  {author} {\bibinfo {author} {\bibfnamefont {R.~F.}\ \bibnamefont
  {Marzke}}\ and\ \bibinfo {author} {\bibfnamefont {J.~A.}\ \bibnamefont
  {Wheeler}},\ }\href@noop {} {\bibfield  {journal} {\bibinfo  {journal}
  {Gravitation and relativity}\ ,\ \bibinfo {pages} {40}} (\bibinfo {year}
  {1964})}\BibitemShut {NoStop}%
\bibitem [{\citenamefont {Romano}\ and\ \citenamefont
  {Cornish}(2017)}]{Romano:2016dpx}%
  \BibitemOpen
  \bibfield  {author} {\bibinfo {author} {\bibfnamefont {J.~D.}\ \bibnamefont
  {Romano}}\ and\ \bibinfo {author} {\bibfnamefont {N.~J.}\ \bibnamefont
  {Cornish}},\ }\href {https://doi.org/10.1007/s41114-017-0004-1} {\bibfield
  {journal} {\bibinfo  {journal} {Living Rev. Rel.}\ }\textbf {\bibinfo
  {volume} {20}},\ \bibinfo {pages} {2} (\bibinfo {year} {2017})},\ \Eprint
  {https://arxiv.org/abs/1608.06889} {arXiv:1608.06889 [gr-qc]} \BibitemShut
  {NoStop}%
\bibitem [{\citenamefont {Estabrook}\ and\ \citenamefont
  {Wahlquist}(1975)}]{Estabrook:1975jtn}%
  \BibitemOpen
  \bibfield  {author} {\bibinfo {author} {\bibfnamefont {F.~B.}\ \bibnamefont
  {Estabrook}}\ and\ \bibinfo {author} {\bibfnamefont {H.~D.}\ \bibnamefont
  {Wahlquist}},\ }\href {https://doi.org/10.1007/BF00762449} {\bibfield
  {journal} {\bibinfo  {journal} {Gen. Rel. Grav.}\ }\textbf {\bibinfo {volume}
  {6}},\ \bibinfo {pages} {439} (\bibinfo {year} {1975})}\BibitemShut {NoStop}%
\bibitem [{\citenamefont {Armstrong}(2006)}]{Armstrong:2006zz}%
  \BibitemOpen
  \bibfield  {author} {\bibinfo {author} {\bibfnamefont {J.~W.}\ \bibnamefont
  {Armstrong}},\ }\href {https://doi.org/10.12942/lrr-2006-1} {\bibfield
  {journal} {\bibinfo  {journal} {Living Rev. Rel.}\ }\textbf {\bibinfo
  {volume} {9}},\ \bibinfo {pages} {1} (\bibinfo {year} {2006})}\BibitemShut
  {NoStop}%
\bibitem [{\citenamefont {Zwick}\ \emph {et~al.}(2025)\citenamefont {Zwick},
  \citenamefont {Soyuer}, \citenamefont {D'Orazio}, \citenamefont {O'Neill},
  \citenamefont {Derdzinski}, \citenamefont {Saha}, \citenamefont {Blas},
  \citenamefont {Jenkins},\ and\ \citenamefont {Kelley}}]{Zwick:2024hag}%
  \BibitemOpen
  \bibfield  {author} {\bibinfo {author} {\bibfnamefont {L.}~\bibnamefont
  {Zwick}}, \bibinfo {author} {\bibfnamefont {D.}~\bibnamefont {Soyuer}},
  \bibinfo {author} {\bibfnamefont {D.~J.}\ \bibnamefont {D'Orazio}}, \bibinfo
  {author} {\bibfnamefont {D.}~\bibnamefont {O'Neill}}, \bibinfo {author}
  {\bibfnamefont {A.}~\bibnamefont {Derdzinski}}, \bibinfo {author}
  {\bibfnamefont {P.}~\bibnamefont {Saha}}, \bibinfo {author} {\bibfnamefont
  {D.}~\bibnamefont {Blas}}, \bibinfo {author} {\bibfnamefont {A.~C.}\
  \bibnamefont {Jenkins}},\ and\ \bibinfo {author} {\bibfnamefont {L.~Z.}\
  \bibnamefont {Kelley}},\ }\href {https://doi.org/10.1103/qp2h-y4b2}
  {\bibfield  {journal} {\bibinfo  {journal} {Phys. Rev. D}\ }\textbf {\bibinfo
  {volume} {112}},\ \bibinfo {pages} {083029} (\bibinfo {year} {2025})},\
  \Eprint {https://arxiv.org/abs/2406.02306} {arXiv:2406.02306 [astro-ph.HE]}
  \BibitemShut {NoStop}%
\bibitem [{\citenamefont {Schmidt}\ and\ \citenamefont
  {Jeong}(2012)}]{Schmidt:2012ne}%
  \BibitemOpen
  \bibfield  {author} {\bibinfo {author} {\bibfnamefont {F.}~\bibnamefont
  {Schmidt}}\ and\ \bibinfo {author} {\bibfnamefont {D.}~\bibnamefont
  {Jeong}},\ }\href {https://doi.org/10.1103/PhysRevD.86.083527} {\bibfield
  {journal} {\bibinfo  {journal} {Phys. Rev. D}\ }\textbf {\bibinfo {volume}
  {86}},\ \bibinfo {pages} {083527} (\bibinfo {year} {2012})},\ \Eprint
  {https://arxiv.org/abs/1204.3625} {arXiv:1204.3625 [astro-ph.CO]}
  \BibitemShut {NoStop}%
\bibitem [{\citenamefont {Jeong}\ and\ \citenamefont
  {Schmidt}(2014)}]{Jeong:2013psa}%
  \BibitemOpen
  \bibfield  {author} {\bibinfo {author} {\bibfnamefont {D.}~\bibnamefont
  {Jeong}}\ and\ \bibinfo {author} {\bibfnamefont {F.}~\bibnamefont
  {Schmidt}},\ }\href {https://doi.org/10.1103/PhysRevD.89.043519} {\bibfield
  {journal} {\bibinfo  {journal} {Phys. Rev. D}\ }\textbf {\bibinfo {volume}
  {89}},\ \bibinfo {pages} {043519} (\bibinfo {year} {2014})},\ \Eprint
  {https://arxiv.org/abs/1305.1299} {arXiv:1305.1299 [astro-ph.CO]}
  \BibitemShut {NoStop}%
\bibitem [{\citenamefont {Jeong}\ and\ \citenamefont
  {Schmidt}(2015)}]{Jeong:2014ufa}%
  \BibitemOpen
  \bibfield  {author} {\bibinfo {author} {\bibfnamefont {D.}~\bibnamefont
  {Jeong}}\ and\ \bibinfo {author} {\bibfnamefont {F.}~\bibnamefont
  {Schmidt}},\ }\href {https://doi.org/10.1088/0264-9381/32/4/044001}
  {\bibfield  {journal} {\bibinfo  {journal} {Class. Quant. Grav.}\ }\textbf
  {\bibinfo {volume} {32}},\ \bibinfo {pages} {044001} (\bibinfo {year}
  {2015})},\ \Eprint {https://arxiv.org/abs/1407.7979} {arXiv:1407.7979
  [astro-ph.CO]} \BibitemShut {NoStop}%
\bibitem [{\citenamefont {Dai}\ \emph {et~al.}(2015)\citenamefont {Dai},
  \citenamefont {Pajer},\ and\ \citenamefont {Schmidt}}]{Dai:2015rda}%
  \BibitemOpen
  \bibfield  {author} {\bibinfo {author} {\bibfnamefont {L.}~\bibnamefont
  {Dai}}, \bibinfo {author} {\bibfnamefont {E.}~\bibnamefont {Pajer}},\ and\
  \bibinfo {author} {\bibfnamefont {F.}~\bibnamefont {Schmidt}},\ }\href
  {https://doi.org/10.1088/1475-7516/2015/11/043} {\bibfield  {journal}
  {\bibinfo  {journal} {JCAP}\ }\textbf {\bibinfo {volume} {11}},\ \bibinfo
  {pages} {043}},\ \Eprint {https://arxiv.org/abs/1502.02011} {arXiv:1502.02011
  [gr-qc]} \BibitemShut {NoStop}%
\bibitem [{\citenamefont {Brill}(1972)}]{brill_redshift}%
  \BibitemOpen
  \bibfield  {author} {\bibinfo {author} {\bibfnamefont {D.~R.}\ \bibnamefont
  {Brill}},\ }in\ \href@noop {} {\emph {\bibinfo {booktitle} {Methods of Local
  and Global Differential Geometry in General Relativity}}},\ \bibinfo {editor}
  {edited by\ \bibinfo {editor} {\bibfnamefont {D.}~\bibnamefont {Farnsworth}},
  \bibinfo {editor} {\bibfnamefont {J.}~\bibnamefont {Fink}}, \bibinfo {editor}
  {\bibfnamefont {J.}~\bibnamefont {Porter}},\ and\ \bibinfo {editor}
  {\bibfnamefont {A.}~\bibnamefont {Thompson}}}\ (\bibinfo  {publisher}
  {Springer Berlin Heidelberg},\ \bibinfo {address} {Berlin, Heidelberg},\
  \bibinfo {year} {1972})\ pp.\ \bibinfo {pages} {45--47}\BibitemShut {NoStop}%
\bibitem [{\citenamefont {Domènech}\ \emph {et~al.}(2025)\citenamefont
  {Domènech}, \citenamefont {Pi},\ and\ \citenamefont {Wang}}]{InPrep}%
  \BibitemOpen
  \bibfield  {author} {\bibinfo {author} {\bibfnamefont {G.}~\bibnamefont
  {Domènech}}, \bibinfo {author} {\bibfnamefont {S.}~\bibnamefont {Pi}},\ and\
  \bibinfo {author} {\bibfnamefont {A.}~\bibnamefont {Wang}}} (\bibinfo {year}
  {2025}),\ \bibinfo {note} {in preparation}\BibitemShut {NoStop}%
\bibitem [{\citenamefont {Kodama}\ and\ \citenamefont
  {Sasaki}(1984)}]{Kodama:1984ziu}%
  \BibitemOpen
  \bibfield  {author} {\bibinfo {author} {\bibfnamefont {H.}~\bibnamefont
  {Kodama}}\ and\ \bibinfo {author} {\bibfnamefont {M.}~\bibnamefont
  {Sasaki}},\ }\href {https://doi.org/10.1143/PTPS.78.1} {\bibfield  {journal}
  {\bibinfo  {journal} {Prog. Theor. Phys. Suppl.}\ }\textbf {\bibinfo {volume}
  {78}},\ \bibinfo {pages} {1} (\bibinfo {year} {1984})}\BibitemShut {NoStop}%
\bibitem [{\citenamefont {Mukhanov}\ \emph {et~al.}(1992)\citenamefont
  {Mukhanov}, \citenamefont {Feldman},\ and\ \citenamefont
  {Brandenberger}}]{Mukhanov:1990me}%
  \BibitemOpen
  \bibfield  {author} {\bibinfo {author} {\bibfnamefont {V.~F.}\ \bibnamefont
  {Mukhanov}}, \bibinfo {author} {\bibfnamefont {H.~A.}\ \bibnamefont
  {Feldman}},\ and\ \bibinfo {author} {\bibfnamefont {R.~H.}\ \bibnamefont
  {Brandenberger}},\ }\href {https://doi.org/10.1016/0370-1573(92)90044-Z}
  {\bibfield  {journal} {\bibinfo  {journal} {Phys. Rept.}\ }\textbf {\bibinfo
  {volume} {215}},\ \bibinfo {pages} {203} (\bibinfo {year}
  {1992})}\BibitemShut {NoStop}%
\bibitem [{\citenamefont {Malik}\ and\ \citenamefont
  {Wands}(2009)}]{Malik:2008im}%
  \BibitemOpen
  \bibfield  {author} {\bibinfo {author} {\bibfnamefont {K.~A.}\ \bibnamefont
  {Malik}}\ and\ \bibinfo {author} {\bibfnamefont {D.}~\bibnamefont {Wands}},\
  }\href {https://doi.org/10.1016/j.physrep.2009.03.001} {\bibfield  {journal}
  {\bibinfo  {journal} {Phys. Rept.}\ }\textbf {\bibinfo {volume} {475}},\
  \bibinfo {pages} {1} (\bibinfo {year} {2009})},\ \Eprint
  {https://arxiv.org/abs/0809.4944} {arXiv:0809.4944 [astro-ph]} \BibitemShut
  {NoStop}%
\bibitem [{\citenamefont {Bonvin}\ and\ \citenamefont
  {Durrer}(2011)}]{Bonvin:2011bg}%
  \BibitemOpen
  \bibfield  {author} {\bibinfo {author} {\bibfnamefont {C.}~\bibnamefont
  {Bonvin}}\ and\ \bibinfo {author} {\bibfnamefont {R.}~\bibnamefont
  {Durrer}},\ }\href {https://doi.org/10.1103/PhysRevD.84.063505} {\bibfield
  {journal} {\bibinfo  {journal} {Phys. Rev. D}\ }\textbf {\bibinfo {volume}
  {84}},\ \bibinfo {pages} {063505} (\bibinfo {year} {2011})},\ \Eprint
  {https://arxiv.org/abs/1105.5280} {arXiv:1105.5280 [astro-ph.CO]}
  \BibitemShut {NoStop}%
\bibitem [{\citenamefont {Durrer}(1994)}]{Durrer:1993tti}%
  \BibitemOpen
  \bibfield  {author} {\bibinfo {author} {\bibfnamefont {R.}~\bibnamefont
  {Durrer}},\ }\href@noop {} {\bibfield  {journal} {\bibinfo  {journal} {Fund.
  Cosmic Phys.}\ }\textbf {\bibinfo {volume} {15}},\ \bibinfo {pages} {209}
  (\bibinfo {year} {1994})},\ \Eprint {https://arxiv.org/abs/astro-ph/9311041}
  {arXiv:astro-ph/9311041} \BibitemShut {NoStop}%
\bibitem [{\citenamefont {Di~Dio}\ and\ \citenamefont
  {Durrer}(2012)}]{DiDio:2012bu}%
  \BibitemOpen
  \bibfield  {author} {\bibinfo {author} {\bibfnamefont {E.}~\bibnamefont
  {Di~Dio}}\ and\ \bibinfo {author} {\bibfnamefont {R.}~\bibnamefont
  {Durrer}},\ }\href {https://doi.org/10.1103/PhysRevD.86.023510} {\bibfield
  {journal} {\bibinfo  {journal} {Phys. Rev. D}\ }\textbf {\bibinfo {volume}
  {86}},\ \bibinfo {pages} {023510} (\bibinfo {year} {2012})},\ \Eprint
  {https://arxiv.org/abs/1205.3366} {arXiv:1205.3366 [astro-ph.CO]}
  \BibitemShut {NoStop}%
\bibitem [{\citenamefont {Gasperini}\ \emph {et~al.}(2011)\citenamefont
  {Gasperini}, \citenamefont {Marozzi}, \citenamefont {Nugier},\ and\
  \citenamefont {Veneziano}}]{Gasperini:2011us}%
  \BibitemOpen
  \bibfield  {author} {\bibinfo {author} {\bibfnamefont {M.}~\bibnamefont
  {Gasperini}}, \bibinfo {author} {\bibfnamefont {G.}~\bibnamefont {Marozzi}},
  \bibinfo {author} {\bibfnamefont {F.}~\bibnamefont {Nugier}},\ and\ \bibinfo
  {author} {\bibfnamefont {G.}~\bibnamefont {Veneziano}},\ }\href
  {https://doi.org/10.1088/1475-7516/2011/07/008} {\bibfield  {journal}
  {\bibinfo  {journal} {JCAP}\ }\textbf {\bibinfo {volume} {07}},\ \bibinfo
  {pages} {008}},\ \Eprint {https://arxiv.org/abs/1104.1167} {arXiv:1104.1167
  [astro-ph.CO]} \BibitemShut {NoStop}%
\bibitem [{\citenamefont {Ben-Dayan}\ \emph {et~al.}(2012)\citenamefont
  {Ben-Dayan}, \citenamefont {Marozzi}, \citenamefont {Nugier},\ and\
  \citenamefont {Veneziano}}]{Ben-Dayan:2012lcv}%
  \BibitemOpen
  \bibfield  {author} {\bibinfo {author} {\bibfnamefont {I.}~\bibnamefont
  {Ben-Dayan}}, \bibinfo {author} {\bibfnamefont {G.}~\bibnamefont {Marozzi}},
  \bibinfo {author} {\bibfnamefont {F.}~\bibnamefont {Nugier}},\ and\ \bibinfo
  {author} {\bibfnamefont {G.}~\bibnamefont {Veneziano}},\ }\href
  {https://doi.org/10.1088/1475-7516/2012/11/045} {\bibfield  {journal}
  {\bibinfo  {journal} {JCAP}\ }\textbf {\bibinfo {volume} {11}},\ \bibinfo
  {pages} {045}},\ \Eprint {https://arxiv.org/abs/1209.4326} {arXiv:1209.4326
  [astro-ph.CO]} \BibitemShut {NoStop}%
\bibitem [{\citenamefont {Bertacca}\ \emph {et~al.}(2014)\citenamefont
  {Bertacca}, \citenamefont {Maartens},\ and\ \citenamefont
  {Clarkson}}]{Bertacca:2014wga}%
  \BibitemOpen
  \bibfield  {author} {\bibinfo {author} {\bibfnamefont {D.}~\bibnamefont
  {Bertacca}}, \bibinfo {author} {\bibfnamefont {R.}~\bibnamefont {Maartens}},\
  and\ \bibinfo {author} {\bibfnamefont {C.}~\bibnamefont {Clarkson}},\ }\href
  {https://doi.org/10.1088/1475-7516/2014/11/013} {\bibfield  {journal}
  {\bibinfo  {journal} {JCAP}\ }\textbf {\bibinfo {volume} {11}},\ \bibinfo
  {pages} {013}},\ \Eprint {https://arxiv.org/abs/1406.0319} {arXiv:1406.0319
  [astro-ph.CO]} \BibitemShut {NoStop}%
\bibitem [{\citenamefont {Marozzi}(2015)}]{Marozzi:2014kua}%
  \BibitemOpen
  \bibfield  {author} {\bibinfo {author} {\bibfnamefont {G.}~\bibnamefont
  {Marozzi}},\ }\href {https://doi.org/10.1088/0264-9381/32/4/045004}
  {\bibfield  {journal} {\bibinfo  {journal} {Class. Quant. Grav.}\ }\textbf
  {\bibinfo {volume} {32}},\ \bibinfo {pages} {045004} (\bibinfo {year}
  {2015})},\ \bibinfo {note} {[Erratum: Class.Quant.Grav. 32, 179501 (2015)]},\
  \Eprint {https://arxiv.org/abs/1406.1135} {arXiv:1406.1135 [astro-ph.CO]}
  \BibitemShut {NoStop}%
\bibitem [{\citenamefont {Yoo}\ and\ \citenamefont
  {Durrer}(2017)}]{Yoo:2017svj}%
  \BibitemOpen
  \bibfield  {author} {\bibinfo {author} {\bibfnamefont {J.}~\bibnamefont
  {Yoo}}\ and\ \bibinfo {author} {\bibfnamefont {R.}~\bibnamefont {Durrer}},\
  }\href {https://doi.org/10.1088/1475-7516/2017/09/016} {\bibfield  {journal}
  {\bibinfo  {journal} {JCAP}\ }\textbf {\bibinfo {volume} {09}},\ \bibinfo
  {pages} {016}},\ \Eprint {https://arxiv.org/abs/1705.05839} {arXiv:1705.05839
  [astro-ph.CO]} \BibitemShut {NoStop}%
\bibitem [{\citenamefont {Fuentes}\ \emph {et~al.}(2021)\citenamefont
  {Fuentes}, \citenamefont {Hidalgo},\ and\ \citenamefont
  {Malik}}]{Fuentes:2019nel}%
  \BibitemOpen
  \bibfield  {author} {\bibinfo {author} {\bibfnamefont {J.~L.}\ \bibnamefont
  {Fuentes}}, \bibinfo {author} {\bibfnamefont {J.~C.}\ \bibnamefont
  {Hidalgo}},\ and\ \bibinfo {author} {\bibfnamefont {K.~A.}\ \bibnamefont
  {Malik}},\ }\href {https://doi.org/10.1088/1361-6382/abd95c} {\bibfield
  {journal} {\bibinfo  {journal} {Class. Quant. Grav.}\ }\textbf {\bibinfo
  {volume} {38}},\ \bibinfo {pages} {065014} (\bibinfo {year} {2021})},\
  \Eprint {https://arxiv.org/abs/1908.08400} {arXiv:1908.08400 [astro-ph.CO]}
  \BibitemShut {NoStop}%
\bibitem [{\citenamefont {Magi}\ and\ \citenamefont
  {Yoo}(2022)}]{Magi:2022nfy}%
  \BibitemOpen
  \bibfield  {author} {\bibinfo {author} {\bibfnamefont {M.}~\bibnamefont
  {Magi}}\ and\ \bibinfo {author} {\bibfnamefont {J.}~\bibnamefont {Yoo}},\
  }\href {https://doi.org/10.1088/1475-7516/2022/09/071} {\bibfield  {journal}
  {\bibinfo  {journal} {JCAP}\ }\textbf {\bibinfo {volume} {09}},\ \bibinfo
  {pages} {071}},\ \Eprint {https://arxiv.org/abs/2204.01751} {arXiv:2204.01751
  [astro-ph.CO]} \BibitemShut {NoStop}%
\bibitem [{\citenamefont {B{\'e}chaz}\ \emph {et~al.}(2025)\citenamefont
  {B{\'e}chaz}, \citenamefont {Fanizza}, \citenamefont {Marozzi},\ and\
  \citenamefont {Silva}}]{Bechaz:2025ojy}%
  \BibitemOpen
  \bibfield  {author} {\bibinfo {author} {\bibfnamefont {P.}~\bibnamefont
  {B{\'e}chaz}}, \bibinfo {author} {\bibfnamefont {G.}~\bibnamefont {Fanizza}},
  \bibinfo {author} {\bibfnamefont {G.}~\bibnamefont {Marozzi}},\ and\ \bibinfo
  {author} {\bibfnamefont {M.~R.~M.}\ \bibnamefont {Silva}},\ }\href@noop {} {\
   (\bibinfo {year} {2025})},\ \Eprint {https://arxiv.org/abs/2510.25690}
  {arXiv:2510.25690 [astro-ph.CO]} \BibitemShut {NoStop}%
\bibitem [{\citenamefont {Romano}\ and\ \citenamefont
  {Allen}(2024)}]{Romano:2023zhb}%
  \BibitemOpen
  \bibfield  {author} {\bibinfo {author} {\bibfnamefont {J.~D.}\ \bibnamefont
  {Romano}}\ and\ \bibinfo {author} {\bibfnamefont {B.}~\bibnamefont {Allen}},\
  }\href {https://doi.org/10.1088/1361-6382/ad4c4c} {\bibfield  {journal}
  {\bibinfo  {journal} {Class. Quant. Grav.}\ }\textbf {\bibinfo {volume}
  {41}},\ \bibinfo {pages} {175008} (\bibinfo {year} {2024})},\ \Eprint
  {https://arxiv.org/abs/2308.05847} {arXiv:2308.05847 [gr-qc]} \BibitemShut
  {NoStop}%
\bibitem [{\citenamefont {Hellings}\ and\ \citenamefont
  {Downs}(1983)}]{Hellings:1983fr}%
  \BibitemOpen
  \bibfield  {author} {\bibinfo {author} {\bibfnamefont {R.~w.}\ \bibnamefont
  {Hellings}}\ and\ \bibinfo {author} {\bibfnamefont {G.~s.}\ \bibnamefont
  {Downs}},\ }\href {https://doi.org/10.1086/183954} {\bibfield  {journal}
  {\bibinfo  {journal} {Astrophys. J. Lett.}\ }\textbf {\bibinfo {volume}
  {265}},\ \bibinfo {pages} {L39} (\bibinfo {year} {1983})}\BibitemShut
  {NoStop}%
\bibitem [{\citenamefont {Mollerach}\ \emph {et~al.}(2004)\citenamefont
  {Mollerach}, \citenamefont {Harari},\ and\ \citenamefont
  {Matarrese}}]{Mollerach:2003nq}%
  \BibitemOpen
  \bibfield  {author} {\bibinfo {author} {\bibfnamefont {S.}~\bibnamefont
  {Mollerach}}, \bibinfo {author} {\bibfnamefont {D.}~\bibnamefont {Harari}},\
  and\ \bibinfo {author} {\bibfnamefont {S.}~\bibnamefont {Matarrese}},\ }\href
  {https://doi.org/10.1103/PhysRevD.69.063002} {\bibfield  {journal} {\bibinfo
  {journal} {Phys. Rev. D}\ }\textbf {\bibinfo {volume} {69}},\ \bibinfo
  {pages} {063002} (\bibinfo {year} {2004})},\ \Eprint
  {https://arxiv.org/abs/astro-ph/0310711} {arXiv:astro-ph/0310711}
  \BibitemShut {NoStop}%
\bibitem [{\citenamefont {Assadullahi}\ and\ \citenamefont
  {Wands}(2009)}]{Assadullahi:2009nf}%
  \BibitemOpen
  \bibfield  {author} {\bibinfo {author} {\bibfnamefont {H.}~\bibnamefont
  {Assadullahi}}\ and\ \bibinfo {author} {\bibfnamefont {D.}~\bibnamefont
  {Wands}},\ }\href {https://doi.org/10.1103/PhysRevD.79.083511} {\bibfield
  {journal} {\bibinfo  {journal} {Phys. Rev. D}\ }\textbf {\bibinfo {volume}
  {79}},\ \bibinfo {pages} {083511} (\bibinfo {year} {2009})},\ \Eprint
  {https://arxiv.org/abs/0901.0989} {arXiv:0901.0989 [astro-ph.CO]}
  \BibitemShut {NoStop}%
\bibitem [{\citenamefont {Papanikolaou}\ \emph {et~al.}(2021)\citenamefont
  {Papanikolaou}, \citenamefont {Vennin},\ and\ \citenamefont
  {Langlois}}]{Papanikolaou:2020qtd}%
  \BibitemOpen
  \bibfield  {author} {\bibinfo {author} {\bibfnamefont {T.}~\bibnamefont
  {Papanikolaou}}, \bibinfo {author} {\bibfnamefont {V.}~\bibnamefont
  {Vennin}},\ and\ \bibinfo {author} {\bibfnamefont {D.}~\bibnamefont
  {Langlois}},\ }\href {https://doi.org/10.1088/1475-7516/2021/03/053}
  {\bibfield  {journal} {\bibinfo  {journal} {JCAP}\ }\textbf {\bibinfo
  {volume} {03}},\ \bibinfo {pages} {053}},\ \Eprint
  {https://arxiv.org/abs/2010.11573} {arXiv:2010.11573 [astro-ph.CO]}
  \BibitemShut {NoStop}%
\bibitem [{\citenamefont {Sipp}\ and\ \citenamefont
  {Schaefer}(2023)}]{Sipp:2022kmb}%
  \BibitemOpen
  \bibfield  {author} {\bibinfo {author} {\bibfnamefont {M.}~\bibnamefont
  {Sipp}}\ and\ \bibinfo {author} {\bibfnamefont {B.~M.}\ \bibnamefont
  {Schaefer}},\ }\href {https://doi.org/10.1103/PhysRevD.107.063538} {\bibfield
   {journal} {\bibinfo  {journal} {Phys. Rev. D}\ }\textbf {\bibinfo {volume}
  {107}},\ \bibinfo {pages} {063538} (\bibinfo {year} {2023})},\ \Eprint
  {https://arxiv.org/abs/2212.01190} {arXiv:2212.01190 [astro-ph.CO]}
  \BibitemShut {NoStop}%
\bibitem [{\citenamefont {Inomata}\ \emph
  {et~al.}(2019{\natexlab{a}})\citenamefont {Inomata}, \citenamefont {Kohri},
  \citenamefont {Nakama},\ and\ \citenamefont {Terada}}]{Inomata:2019ivs}%
  \BibitemOpen
  \bibfield  {author} {\bibinfo {author} {\bibfnamefont {K.}~\bibnamefont
  {Inomata}}, \bibinfo {author} {\bibfnamefont {K.}~\bibnamefont {Kohri}},
  \bibinfo {author} {\bibfnamefont {T.}~\bibnamefont {Nakama}},\ and\ \bibinfo
  {author} {\bibfnamefont {T.}~\bibnamefont {Terada}},\ }\href
  {https://doi.org/10.1103/PhysRevD.108.049901} {\bibfield  {journal} {\bibinfo
   {journal} {Phys. Rev. D}\ }\textbf {\bibinfo {volume} {100}},\ \bibinfo
  {pages} {043532} (\bibinfo {year} {2019}{\natexlab{a}})},\ \bibinfo {note}
  {[Erratum: Phys.Rev.D 108, 049901 (2023)]},\ \Eprint
  {https://arxiv.org/abs/1904.12879} {arXiv:1904.12879 [astro-ph.CO]}
  \BibitemShut {NoStop}%
\bibitem [{\citenamefont {Inomata}\ \emph
  {et~al.}(2019{\natexlab{b}})\citenamefont {Inomata}, \citenamefont {Kohri},
  \citenamefont {Nakama},\ and\ \citenamefont {Terada}}]{Inomata:2019zqy}%
  \BibitemOpen
  \bibfield  {author} {\bibinfo {author} {\bibfnamefont {K.}~\bibnamefont
  {Inomata}}, \bibinfo {author} {\bibfnamefont {K.}~\bibnamefont {Kohri}},
  \bibinfo {author} {\bibfnamefont {T.}~\bibnamefont {Nakama}},\ and\ \bibinfo
  {author} {\bibfnamefont {T.}~\bibnamefont {Terada}},\ }\href
  {https://doi.org/10.1088/1475-7516/2019/10/071} {\bibfield  {journal}
  {\bibinfo  {journal} {JCAP}\ }\textbf {\bibinfo {volume} {10}},\ \bibinfo
  {pages} {071}},\ \bibinfo {note} {[Erratum: JCAP 08, E01 (2023)]},\ \Eprint
  {https://arxiv.org/abs/1904.12878} {arXiv:1904.12878 [astro-ph.CO]}
  \BibitemShut {NoStop}%
\bibitem [{\citenamefont {Naruko}\ \emph {et~al.}(2013)\citenamefont {Naruko},
  \citenamefont {Pitrou}, \citenamefont {Koyama},\ and\ \citenamefont
  {Sasaki}}]{Naruko:2013aaa}%
  \BibitemOpen
  \bibfield  {author} {\bibinfo {author} {\bibfnamefont {A.}~\bibnamefont
  {Naruko}}, \bibinfo {author} {\bibfnamefont {C.}~\bibnamefont {Pitrou}},
  \bibinfo {author} {\bibfnamefont {K.}~\bibnamefont {Koyama}},\ and\ \bibinfo
  {author} {\bibfnamefont {M.}~\bibnamefont {Sasaki}},\ }\href
  {https://doi.org/10.1088/0264-9381/30/16/165008} {\bibfield  {journal}
  {\bibinfo  {journal} {Class. Quant. Grav.}\ }\textbf {\bibinfo {volume}
  {30}},\ \bibinfo {pages} {165008} (\bibinfo {year} {2013})},\ \Eprint
  {https://arxiv.org/abs/1304.6929} {arXiv:1304.6929 [astro-ph.CO]}
  \BibitemShut {NoStop}%
\bibitem [{\citenamefont {Saito}\ \emph {et~al.}(2014)\citenamefont {Saito},
  \citenamefont {Naruko}, \citenamefont {Hiramatsu},\ and\ \citenamefont
  {Sasaki}}]{Saito:2014bxa}%
  \BibitemOpen
  \bibfield  {author} {\bibinfo {author} {\bibfnamefont {R.}~\bibnamefont
  {Saito}}, \bibinfo {author} {\bibfnamefont {A.}~\bibnamefont {Naruko}},
  \bibinfo {author} {\bibfnamefont {T.}~\bibnamefont {Hiramatsu}},\ and\
  \bibinfo {author} {\bibfnamefont {M.}~\bibnamefont {Sasaki}},\ }\href
  {https://doi.org/10.1088/1475-7516/2014/10/051} {\bibfield  {journal}
  {\bibinfo  {journal} {JCAP}\ }\textbf {\bibinfo {volume} {10}},\ \bibinfo
  {pages} {051}},\ \Eprint {https://arxiv.org/abs/1409.2464} {arXiv:1409.2464
  [astro-ph.CO]} \BibitemShut {NoStop}%
\bibitem [{\citenamefont {Namikawa}\ \emph {et~al.}(2021)\citenamefont
  {Namikawa}, \citenamefont {Naruko}, \citenamefont {Saito}, \citenamefont
  {Taruya},\ and\ \citenamefont {Yamauchi}}]{Namikawa:2021obu}%
  \BibitemOpen
  \bibfield  {author} {\bibinfo {author} {\bibfnamefont {T.}~\bibnamefont
  {Namikawa}}, \bibinfo {author} {\bibfnamefont {A.}~\bibnamefont {Naruko}},
  \bibinfo {author} {\bibfnamefont {R.}~\bibnamefont {Saito}}, \bibinfo
  {author} {\bibfnamefont {A.}~\bibnamefont {Taruya}},\ and\ \bibinfo {author}
  {\bibfnamefont {D.}~\bibnamefont {Yamauchi}},\ }\href
  {https://doi.org/10.1088/1475-7516/2021/10/029} {\bibfield  {journal}
  {\bibinfo  {journal} {JCAP}\ }\textbf {\bibinfo {volume} {10}},\ \bibinfo
  {pages} {029}},\ \Eprint {https://arxiv.org/abs/2103.10639} {arXiv:2103.10639
  [astro-ph.CO]} \BibitemShut {NoStop}%
\bibitem [{\citenamefont {Gurian}\ \emph {et~al.}(2021)\citenamefont {Gurian},
  \citenamefont {Jeong}, \citenamefont {Hwang},\ and\ \citenamefont
  {Noh}}]{Gurian:2021rfv}%
  \BibitemOpen
  \bibfield  {author} {\bibinfo {author} {\bibfnamefont {J.}~\bibnamefont
  {Gurian}}, \bibinfo {author} {\bibfnamefont {D.}~\bibnamefont {Jeong}},
  \bibinfo {author} {\bibfnamefont {J.-c.}\ \bibnamefont {Hwang}},\ and\
  \bibinfo {author} {\bibfnamefont {H.}~\bibnamefont {Noh}},\ }\href
  {https://doi.org/10.1103/PhysRevD.104.083534} {\bibfield  {journal} {\bibinfo
   {journal} {Phys. Rev. D}\ }\textbf {\bibinfo {volume} {104}},\ \bibinfo
  {pages} {083534} (\bibinfo {year} {2021})},\ \Eprint
  {https://arxiv.org/abs/2104.03330} {arXiv:2104.03330 [astro-ph.CO]}
  \BibitemShut {NoStop}%
\bibitem [{\citenamefont {Cai}\ \emph {et~al.}(2022)\citenamefont {Cai},
  \citenamefont {Yang},\ and\ \citenamefont {Zhao}}]{Cai:2021jbi}%
  \BibitemOpen
  \bibfield  {author} {\bibinfo {author} {\bibfnamefont {R.-G.}\ \bibnamefont
  {Cai}}, \bibinfo {author} {\bibfnamefont {X.-Y.}\ \bibnamefont {Yang}},\ and\
  \bibinfo {author} {\bibfnamefont {L.}~\bibnamefont {Zhao}},\ }\href
  {https://doi.org/10.1007/s10714-022-02972-x} {\bibfield  {journal} {\bibinfo
  {journal} {Gen. Rel. Grav.}\ }\textbf {\bibinfo {volume} {54}},\ \bibinfo
  {pages} {89} (\bibinfo {year} {2022})},\ \Eprint
  {https://arxiv.org/abs/2109.06864} {arXiv:2109.06864 [gr-qc]} \BibitemShut
  {NoStop}%
\bibitem [{\citenamefont {Favata}(2010)}]{Favata:2010zu}%
  \BibitemOpen
  \bibfield  {author} {\bibinfo {author} {\bibfnamefont {M.}~\bibnamefont
  {Favata}},\ }\href {https://doi.org/10.1088/0264-9381/27/8/084036} {\bibfield
   {journal} {\bibinfo  {journal} {Class. Quant. Grav.}\ }\textbf {\bibinfo
  {volume} {27}},\ \bibinfo {pages} {084036} (\bibinfo {year} {2010})},\
  \Eprint {https://arxiv.org/abs/1003.3486} {arXiv:1003.3486 [gr-qc]}
  \BibitemShut {NoStop}%
\bibitem [{\citenamefont {Harte}\ \emph {et~al.}(2025)\citenamefont {Harte},
  \citenamefont {Mieling}, \citenamefont {Oancea},\ and\ \citenamefont
  {Steininger}}]{Harte:2024mwj}%
  \BibitemOpen
  \bibfield  {author} {\bibinfo {author} {\bibfnamefont {A.~I.}\ \bibnamefont
  {Harte}}, \bibinfo {author} {\bibfnamefont {T.~B.}\ \bibnamefont {Mieling}},
  \bibinfo {author} {\bibfnamefont {M.~A.}\ \bibnamefont {Oancea}},\ and\
  \bibinfo {author} {\bibfnamefont {E.}~\bibnamefont {Steininger}},\ }\href
  {https://doi.org/10.1103/PhysRevD.111.024034} {\bibfield  {journal} {\bibinfo
   {journal} {Phys. Rev. D}\ }\textbf {\bibinfo {volume} {111}},\ \bibinfo
  {pages} {024034} (\bibinfo {year} {2025})},\ \Eprint
  {https://arxiv.org/abs/2407.00174} {arXiv:2407.00174 [gr-qc]} \BibitemShut
  {NoStop}%
\bibitem [{\citenamefont {Chang}\ \emph {et~al.}(2023)\citenamefont {Chang},
  \citenamefont {Zhang},\ and\ \citenamefont {Zhou}}]{Chang:2022vlv}%
  \BibitemOpen
  \bibfield  {author} {\bibinfo {author} {\bibfnamefont {Z.}~\bibnamefont
  {Chang}}, \bibinfo {author} {\bibfnamefont {X.}~\bibnamefont {Zhang}},\ and\
  \bibinfo {author} {\bibfnamefont {J.-Z.}\ \bibnamefont {Zhou}},\ }\href
  {https://doi.org/10.1103/PhysRevD.107.063510} {\bibfield  {journal} {\bibinfo
   {journal} {Phys. Rev. D}\ }\textbf {\bibinfo {volume} {107}},\ \bibinfo
  {pages} {063510} (\bibinfo {year} {2023})},\ \Eprint
  {https://arxiv.org/abs/2209.07693} {arXiv:2209.07693 [astro-ph.CO]}
  \BibitemShut {NoStop}%
\bibitem [{\citenamefont {Yu}\ and\ \citenamefont {Wang}(2024)}]{Yu:2023lmo}%
  \BibitemOpen
  \bibfield  {author} {\bibinfo {author} {\bibfnamefont {Y.-H.}\ \bibnamefont
  {Yu}}\ and\ \bibinfo {author} {\bibfnamefont {S.}~\bibnamefont {Wang}},\
  }\href {https://doi.org/10.1140/epjc/s10052-024-12937-w} {\bibfield
  {journal} {\bibinfo  {journal} {Eur. Phys. J. C}\ }\textbf {\bibinfo {volume}
  {84}},\ \bibinfo {pages} {555} (\bibinfo {year} {2024})},\ \Eprint
  {https://arxiv.org/abs/2303.03897} {arXiv:2303.03897 [astro-ph.CO]}
  \BibitemShut {NoStop}%
\bibitem [{\citenamefont {Bari}\ \emph {et~al.}(2024)\citenamefont {Bari},
  \citenamefont {Bartolo}, \citenamefont {Dom{\`e}nech},\ and\ \citenamefont
  {Matarrese}}]{Bari:2023rcw}%
  \BibitemOpen
  \bibfield  {author} {\bibinfo {author} {\bibfnamefont {P.}~\bibnamefont
  {Bari}}, \bibinfo {author} {\bibfnamefont {N.}~\bibnamefont {Bartolo}},
  \bibinfo {author} {\bibfnamefont {G.}~\bibnamefont {Dom{\`e}nech}},\ and\
  \bibinfo {author} {\bibfnamefont {S.}~\bibnamefont {Matarrese}},\ }\href
  {https://doi.org/10.1103/PhysRevD.109.023509} {\bibfield  {journal} {\bibinfo
   {journal} {Phys. Rev. D}\ }\textbf {\bibinfo {volume} {109}},\ \bibinfo
  {pages} {023509} (\bibinfo {year} {2024})},\ \Eprint
  {https://arxiv.org/abs/2307.05404} {arXiv:2307.05404 [astro-ph.CO]}
  \BibitemShut {NoStop}%
\bibitem [{\citenamefont {Picard}\ and\ \citenamefont
  {Malik}(2024)}]{Picard:2023sbz}%
  \BibitemOpen
  \bibfield  {author} {\bibinfo {author} {\bibfnamefont {R.}~\bibnamefont
  {Picard}}\ and\ \bibinfo {author} {\bibfnamefont {K.~A.}\ \bibnamefont
  {Malik}},\ }\href {https://doi.org/10.1088/1475-7516/2024/10/010} {\bibfield
  {journal} {\bibinfo  {journal} {JCAP}\ }\textbf {\bibinfo {volume} {10}},\
  \bibinfo {pages} {010}},\ \Eprint {https://arxiv.org/abs/2311.14513}
  {arXiv:2311.14513 [astro-ph.CO]} \BibitemShut {NoStop}%
\bibitem [{\citenamefont {Picard}\ \emph {et~al.}(2025)\citenamefont {Picard},
  \citenamefont {Padilla}, \citenamefont {Malik},\ and\ \citenamefont
  {Mulryne}}]{Picard:2025bwq}%
  \BibitemOpen
  \bibfield  {author} {\bibinfo {author} {\bibfnamefont {R.}~\bibnamefont
  {Picard}}, \bibinfo {author} {\bibfnamefont {L.~E.}\ \bibnamefont {Padilla}},
  \bibinfo {author} {\bibfnamefont {K.~A.}\ \bibnamefont {Malik}},\ and\
  \bibinfo {author} {\bibfnamefont {D.~J.}\ \bibnamefont {Mulryne}},\
  }\href@noop {} {\  (\bibinfo {year} {2025})},\ \Eprint
  {https://arxiv.org/abs/2509.07811} {arXiv:2509.07811 [astro-ph.CO]}
  \BibitemShut {NoStop}%
\bibitem [{\citenamefont {Isaacson}(1968)}]{Isaacson:1968zza}%
  \BibitemOpen
  \bibfield  {author} {\bibinfo {author} {\bibfnamefont {R.~A.}\ \bibnamefont
  {Isaacson}},\ }\href {https://doi.org/10.1103/PhysRev.166.1272} {\bibfield
  {journal} {\bibinfo  {journal} {Phys. Rev.}\ }\textbf {\bibinfo {volume}
  {166}},\ \bibinfo {pages} {1272} (\bibinfo {year} {1968})}\BibitemShut
  {NoStop}%
\bibitem [{\citenamefont {Cai}\ \emph {et~al.}(2021)\citenamefont {Cai},
  \citenamefont {Yang},\ and\ \citenamefont {Zhao}}]{Cai:2021ndu}%
  \BibitemOpen
  \bibfield  {author} {\bibinfo {author} {\bibfnamefont {R.-G.}\ \bibnamefont
  {Cai}}, \bibinfo {author} {\bibfnamefont {X.-Y.}\ \bibnamefont {Yang}},\ and\
  \bibinfo {author} {\bibfnamefont {L.}~\bibnamefont {Zhao}},\ }\href@noop {}
  {\  (\bibinfo {year} {2021})},\ \Eprint {https://arxiv.org/abs/2109.06865}
  {arXiv:2109.06865 [astro-ph.CO]} \BibitemShut {NoStop}%
\end{thebibliography}%
\bibliographystyle{apsrev4-2}

\clearpage
\onecolumngrid

\clearpage\onecolumngrid

\begin{center}
\section{End Matter} \label{app:endmatter}
\end{center}

\twocolumngrid

%%%%%%%%%% Merge with supplemental materials %%%%%%%%%%
\setcounter{equation}{0}
\setcounter{figure}{0}
\setcounter{table}{0}
\setcounter{section}{0}
\setcounter{page}{1}
\makeatletter
\renewcommand{\theequation}{E\arabic{equation}}
\renewcommand{\thefigure}{E\arabic{figure}}
\renewcommand{\theHfigure}{E\arabic{figure}}%
\renewcommand{\thetable}{E\arabic{table}}

Here we provide, for the interested reader, some of the derivation details skipped in the main text.

\medskip
\textbf{Boundary conditions.}---
The expansion of the boundary condition, that is
\begin{equation}
    \xvd{\mu}{}\left(\sum_{l =0} \tauf{l}\right)-\xvn{\mu}{\rm f}=0\,,
\end{equation}
up to second order is given by
\begin{align}
    & {\rm 0th}: && \xvd{(0)\mu}{}(\tauf{0})=\xvn{(0)\mu}{\rm f}\label{eq:bcapp}\,,\\
    & {\rm 1st}: && \pv{(0)\mu}{}(\tauf{0})\,\tauf{1}+\xvd{(1)\mu}{}(\tauf{0})=\xvn{(1)\mu}{\rm f}\label{eq:fcapp}\,,\\
    & {\rm 2nd}: && \pv{(0)\mu}{}(\tauf{0})\,\tauf{2} + \frac{1}{2}\frac{{\rm d}\pv{(0)\mu}{}}{{\rm d}\tau_{\rm r}}\Bigg\vert_{\tauf{0}}\,{\tauf{1}}^2\notag\\
    & && \quad + \pv{(1)\mu}{}(\tauf{0})\,\tauf{1}+\xvd{(2)\mu}{}(\tauf{0})=\xvn{(2)\mu}{\rm f} \label{eq:scapp}\,.
\end{align}
\medskip
\textbf{Integration of null condition.}---
The normalization condition requires that the null geodesic satisfies
\begin{equation}
    \kv{\mu}{}\,g_{\mu\nu}\kv{\nu}{}=0\,.
\end{equation}
This condition allows higher-order terms such as $\kv{(l)\mu}{}\gp{0}{\mu\nu}\kv{(0)\nu}{}$ to be expressed in terms of lower-order quantities, thereby significantly reducing the computational complexity. At first order, the condition gives
\begin{align}
    2\kv{(1)\mu}{}\gp{0}{\mu\nu}\kv{(0)\nu}{}+ \kv{(0)\mu}{}\gp{1}{\mu\nu}\kv{(0)\nu}{} =0\,.
\end{align}
After integration, we obtain
\begin{align}\label{eq:null1}
   \xvn{(1)\mu}{}\gp{0}{\mu\nu}\kv{(0)\nu}{}\Big\vert^{w_{\rm f}}_{w_{\rm i}}&=\int_{w_{\rm i}}^{w_{\rm f}}{\rm d}w \,\kv{(1)\mu}{}\gp{0}{\mu\nu}\kv{(0)\nu}{} \nonumber\\&=-\frac{1}{2}\int_{w_{\rm i}}^{w_{\rm f}}{\rm d}w\,\gp{1}{\mathbin{//}}\,,
\end{align}
where we defined the projected metric perturbation along the background photon trajectory as $\gp{l}{\mathbin{//}}\equiv \kv{(0)\mu}{}\gp{l}{\mu\nu}\,\kv{(0)\nu}{}$. 

Eq.~\eqref{eq:null1} can more clearly be written as
\begin{equation}\label{eq:fns}
    \xvn{(1)0}{\rm f}=\xvn{(1)0}{\rm i}+n_{i}\xvn{(1)i}{}\Big\vert^{w_{\rm f}}_{w_{\rm i}}+\frac{1}{2}\int_{w_{\rm i}}^{w_{\rm f}} {\rm d}w\, g^{(1)}_{\mathbin{//}}\,,
\end{equation}
where, after integration by parts, we have that
\begin{align}\label{eq:propagation1app}
 \frac{1}{2}\int_{w_{\rm i}}^{w_{\rm f}} \hspace{-0.1mm}{\rm d}w g^{(1)}_{\mathbin{//}}=\int_{0}^{L}\hspace{-0.1mm}& {\rm d}\lambda\Big(\frac{1}{2}\,n^{i}n^{j}h_{ij}^{(1)}-\Phi^{(1)}-\Psi^{(1)} \Big)\nonumber\\&
 +\big(\sigma^{(1)}+ n^{i} E^{(1)}_{,i}\Big)\Big\vert^{{\rm rf}}_{{\rm ei}}\,.
 \end{align}
 The Bardeen potentials are given by
\begin{align}
\Phi^{(1)} &= \phi^{(1)}+\frac{a'}{a}\sigma^{(1)}+{\sigma^{(1)}}\,,\\
\Psi^{(1)}&=-\psi^{(1)}+\frac{1}{3}\partial^2 E^{(1)}-\frac{a'}{a}\sigma^{(1)}\,.
\end{align} 

At second order, we have that
\begin{align}
    2\kv{(2)\mu}{}\gp{0}{\mu\nu}&\kv{(0)\nu}{}+\kv{(1)\mu}{}\gp{0}{\mu\nu}\kv{(1)\nu}{}\nonumber\\&+ 2\kv{(1)\mu}{}\left(\gp{1}{\mu\nu}+\gp{0}{\mu\nu,\sigma}\xvn{(1)\sigma}{}\right)\kv{(0)\nu}{}\nonumber\\&+\kv{(0)\mu}{}\left(\gp{2}{\mu\nu}+\gp{1}{\mu\nu,\sigma}\xvn{(1)\sigma}{}\right)\kv{(0)\nu}{}=0\,,
\end{align}
and the integral form reads
\begin{align}
    &\xvn{(2)\mu}{}\gp{0}{\mu\nu}\kv{(0)\nu}{}\Big\vert^{w_{\rm f}}_{w_{\rm i}}=\int_{w_{\rm i}}^{w_{\rm f}}{\rm d}w \,\kv{(2)\mu}{}\gp{0}{\mu\nu}\kv{(0)\nu}{}\nonumber\\&=-\frac{1}{2}\int_{w_{\rm i}}^{w_{\rm f}}{\rm d}w\,\Bigg(\gp{2}{\mathbin{//}}+2\kv{(1)\mu}{}\gp{1}{\mu\nu}\kv{(0)\nu}{} +\kv{(1)\mu}{}\gp{0}{\mu\nu}\kv{(1)\nu}{}\notag\\
    &\qquad\qquad\quad+a^2\kv{(0)\mu}{}\left(\gp{1}{\mu\nu}\Big/a^2\right)_{,\sigma}\xvn{(1)\sigma}{}\kv{(0)\nu}{}\Bigg)\label{eq:sng}.
\end{align}
After several integration by parts, and only restricting ourselves to the terms proportional to $n^in^j$ we find that
\begin{align}\label{eq:sns}
    \xvn{(2)0}{\rm f}\supset& \xvn{(2)0}{\rm i}+ n_{i}\xvn{(2)i}{}\Big\vert^{w_{\rm f}}_{w_i}+n^in^j\Big(\frac{1}{2}\int_{0}^{L} {\rm d}\lambda\, h^{N(2)}_{ij}\notag\\
    &+{E^{(1)}_{{\rm rf},i}}'\Delta X^{(1)}_{\gamma i}+\frac{a_{\rm f}^4}{w_{\rm f}^2}\frac{L}{2}\Delta X^{(1)}_{\gamma i}\Delta X^{(1)}_{\gamma j}\Big),
\end{align} 
where we defined
\begin{align}\label{eq:deltaXi}
\Delta \Xvn{1}{i}{}&\equiv\Xvn{1}{i}{\rm f}-\Xvn{1}{i}{\rm i}\\&=\Xvd{(1)i}{\rm i}-\Xve{(1)i}{\rm i}+\int_{\tau_{\rm ri}}^{\tau_{\rm rf}}\frac{{\rm d} {\tau}_{\rm r}}{a^2({\tau}_{\rm r})}V^{(1)}_{{\rm r},i}\,.
\end{align}
Note that $n_i\Delta \Xvn{1}{i}{}=\Delta L_{\rm i}+n_i\Delta X^{(1)}_{i}$; see the discussion around Eqs.~\eqref{eq:stdt} and \eqref{eq:timedelaysecondline}. In Eq.~\eqref{eq:timedelaysecondline} we did not include $\Delta L_{\rm i}$ because it can be absorbed as a redefinition of $L$.

\medskip
\textbf{Time-like geodesics.}--- The solutions to the first-order time-like geodesics read
\begin{align}
    &\pv{(1)0}{}=-\frac{1}{a}\left(\phi^{(1)}+\frac{a'}{a}\xvd{(1)0}{{\rm i}}\right),\\
    &a\,v_{\rm r}^{(1)}=a_{\rm ri}\,V_{\rm ri}^{(1)}-a{E^{(1)}}' +aV_{\rm ac}^{(1)}\label{eq:ftg}\,,
\end{align}
where we defined
\begin{align}
    a(\tau_{\rm r})V_{\rm ac}^{(1)}(\tau_{\rm r})&=-\int_{\tau_{\rm ri}}^{\tau_{\rm rf}} {\rm d} {\tau}_{\rm r}\, \Phi^{(1)}.
\end{align}

Second-order geodesics do not contribute to the quadrupole. Their solution are given by
\begin{align}
    \left(a \xvd{(2)0}{}\right)&\Big\vert^{\tau_{\rm rf}}_{\tau_{\rm ri}}=-\int_{\tau_{\rm ri}}^{\tau_{\rm rf}} {\rm d}{\tau}_{\rm r} A^{(2)}({\tau}_{\rm r})\notag\\&
    \quad+ \Big(-\frac{1}{2}a' \left(\xvd{(1)0}{}\right)^2-a\phi^{(1)}\xvd{(1)0}{}\notag\\
    &\quad+a\left(\beta^{(1)}+v^{(1)}\right)_{,i}\xvd{(1)i}{}\Big)\Big\vert^{\tau_{\rm rf}}_{\tau_{\rm ri}},\\
    \xvd{(2)i}{}&\Big\vert^{\tau_{\rm r}}_{\tau_{\rm ri}}=\int^{\tau_{\rm rf}}_{\tau_{\rm ri}}{\tau}_{\rm r}\,\pv{(2)i}{},
\end{align}
where 
\begin{equation}
    A^{(2)}\equiv\phi^{(2)}+\frac{1}{2} ({\phi^{(1)}})^2+\frac{1}{2}(v^{(1)}_{,i})^2-\frac{1}{2}(\beta^{(1)}_{,i})^2,
\end{equation}
and the second-order velocity correction reads
\begin{align}
    \Big(a^2 \pv{(2)i}{}\Big)&\Big\vert^{\tau_{\rm rf}}_{\tau_{\rm ri}}= \Big(a v^{(1)}_{,i,\rho}\xvd{(1)\rho}{}+a\phi^{(1)}\beta^{(1)}_{,i}-a\beta^{(2)}_{,i}\notag\\&-2a\left(\psi^{(1)}\delta_{ij}+E^{(1)}_{,i,j}\right)(\beta^{(1)}+v^{(1)})_{,j}
    \notag\\&-a'\xvd{(1)0}{}v^{(1)}_{,i}\Big)\Big\vert^{\tau_{\rm rf}}_{\tau_{\rm ri}}
    -\int_{\tau_{\rm ri}}^{\tau_{\rm rf}} {\rm d}{\tau}_{\rm r}\, A^{(2)}_{,i}({\tau}_{\rm r}).
\end{align}

\medskip
\textbf{Second-order time delay.}--- From Eq.~\eqref{eq:scapp} that the second-order correction to the reception time is given by
\begin{align}\label{eq:std}
    \tauf{2}&=\frac{1}{\pv{(0)0}{}(\tauf{0})}\Bigg(\xvn{(2)0}{\rm f}-\xvd{(2)0}{}(\tauf{0})\notag\\
    &\quad\quad-\pv{(1)0}{}(\tauf{0})\tauf{1}-\frac{1}{2}\frac{{\rm d} \pv{(0)0}{}}{{\rm d}\tau_{\rm r}}\Bigg\vert_{\tauf{0}}{\tauf{1}}^2\Bigg)\,.
\end{align}
We now use \eqref{eq:sns} for $\xvn{(2)0}{\rm f}$ supplemented by the boundary conditions, which are $\xvn{(2)\mu}{\rm i}=\xvd{(2)\mu}{}(\tau_{\rm Ei})$ and $\xvn{(2)i}{\rm f}=\xvd{(2)i}{}(\tauf{0})+ \pv{(1)i}{}(\tauf{0})\,\tauf{1}$. 

We find that the second-order correction to the receptor's geodesics does not contribute to the quadrupole. Thus, the remaining term that contributes to the quadrupole is the correction to the space separation due to $\pv{(1)i}{}(\tauf{0})\,\tauf{1}$ and the last term in Eq.~\eqref{eq:sns}. Now, substituting Eq.\eqref{eq:sns} into Eq.\eqref{eq:std}, we obtain the total quadrupole correction to the detection time at second order, that is
\begin{align}\label{eq:stdtapp}
    \tauf{2}&\supset a_{\rm rf}\,n^in^j\Bigg(
    \frac{1}{2}\int_0^{L}{\rm d}\lambda \, h^{N(2)}_{ij}+V^{(1)}_{{\rm rf},i}\Delta \Xvn{1}{j}{}\notag\\
    &+\left(\frac{a'_{\rm rf}}{2a_{\rm rf}^2}-\frac{a_{\rm rf}^4}{2w_{\rm f}^2}L\right)\Delta \Xvn{1}{i}{}\Delta \Xvn{1}{j}{}\Bigg),
\end{align}
which is manifestly gauge invariant. \\

\medskip
\textbf{Useful Formulae for ``Multipole'' Separation.}--- We discussed in the main text that the quadrupole contribution, according to the dependence of the propagation direction $n^i$, represents the GW effect in the first-order case. In the integration of Eq.~\eqref{eq:sng}, many terms can be neglected if we focus on the GW contribution, which also corresponds to the highest multipole effect at second order. For instance, the trace part vanishes because it annihilates two directional indices, while the gradient terms can be integrated by parts as
\begin{equation}
\int_{w_{\rm i}}^{w_{\rm f}} {\rm d}{w}\,n^i F_{,i}/a^2
= F\Big\vert^{w_{\rm f}}_{w_{\rm i}} - \int_{w_{\rm i}}^{w_{\rm f}} {\rm d}{w}\,F'/a^2,
\end{equation}
where $F$ denotes an arbitrary function. Even for the first-order null-like geodesic solution $\kv{(1)\mu}{}$ in Eq.~\eqref{eq:sng}, we can only retain the highest multipole term (i.e., the terms containing $n^i$).

\end{document}